\documentclass[11pt]{article}
\usepackage{lettrine}
\usepackage{cite}
\usepackage[svgnames]{xcolor}
\usepackage{soul}
\usepackage{color}
\usepackage[breakable]{tcolorbox}
\usepackage{amssymb}
\usepackage{bm}
\usepackage{parskip} 
\usepackage{iftex}
\usepackage{framed}
\usepackage{tikz}
\usetikzlibrary{calc}
\ifPDFTeX
    \usepackage[T1]{fontenc}
    \usepackage{mathpazo}
\else
    \usepackage{fontspec}
\fi
\usepackage{fancyhdr}
\usepackage{caption}
\usepackage{subcaption}
\usepackage{float}
\usepackage{xcolor} 
\usepackage{enumerate} 
\usepackage{geometry} 
\usepackage{amssymb} 
\usepackage{textcomp} 
\usepackage{amsmath, amssymb, amsfonts}
\usepackage{graphicx}
\usepackage{listings}
\usepackage[hyperfootnotes=false]{hyperref}
\usepackage{color}
\usepackage{booktabs}   
\usepackage{makecell} 
\definecolor{gray}{rgb}{0.5,0.5,0.5}
\AtBeginDocument{%
}
\usepackage{upquote} 
\usepackage{eurosym} 
\usepackage[mathletters]{ucs} 
\usepackage{fancyvrb} 
\usepackage{grffile} 
\makeatletter
\@ifpackagelater{grffile}{2019/11/01}
{}
{
  \def\Gread@@xetex#1{%
    \IfFileExists{"\Gin@base".bb}%
    {\Gread@eps{\Gin@base.bb}}%
    {\Gread@@xetex@aux#1}%
  }
}
\makeatother
\usepackage[Export]{adjustbox} 
\adjustboxset{max size={0.9\linewidth}{0.9\paperheight}}

\usepackage{titling}
\usepackage{longtable} 
\usepackage{booktabs}  
\usepackage[inline]{enumitem} 
\usepackage[normalem]{ulem} 
\usepackage{mathrsfs}

\usepackage{authblk}
\usepackage{chngcntr}                  

\definecolor{urlcolor}{rgb}{0,.145,.698}
\definecolor{linkcolor}{rgb}{.71,0.21,0.01}
\definecolor{citecolor}{rgb}{.12,.54,.11}
\hypersetup{
    breaklinks=true,
    colorlinks=true,
    urlcolor=urlcolor,
    linkcolor=linkcolor,
    citecolor=citecolor,
}

\definecolor{ansi-black}{HTML}{3E424D}
\definecolor{ansi-black-intense}{HTML}{282C36}
\definecolor{ansi-red}{HTML}{E75C58}
\definecolor{ansi-red-intense}{HTML}{B22B31}
\definecolor{ansi-green}{HTML}{00A250}
\definecolor{ansi-green-intense}{HTML}{007427}
\definecolor{ansi-yellow}{HTML}{DDB62B}
\definecolor{ansi-yellow-intense}{HTML}{B27D12}
\definecolor{ansi-blue}{HTML}{208FFB}
\definecolor{ansi-blue-intense}{HTML}{0065CA}
\definecolor{ansi-magenta}{HTML}{D160C4}
\definecolor{ansi-magenta-intense}{HTML}{A03196}
\definecolor{ansi-cyan}{HTML}{60C6C8}
\definecolor{ansi-cyan-intense}{HTML}{258F8F}
\definecolor{ansi-white}{HTML}{C5C1B4}
\definecolor{ansi-white-intense}{HTML}{A1A6B2}
\definecolor{ansi-default-inverse-fg}{HTML}{FFFFFF}
\definecolor{ansi-default-inverse-bg}{HTML}{000000}
\definecolor{outerrorbackground}{HTML}{FFDFDF}
\definecolor{darkgreen}{rgb}{0.0, 0.5, 0.0}

\DefineVerbatimEnvironment{Highlighting}{Verbatim}{commandchars=\\\{\}}

\makeatletter
\newcommand{\boxspacing}{\kern\kvtcb@left@rule\kern\kvtcb@boxsep}
\makeatother

\title{A convolutional neural network surrogate for hierarchical homogenization: fast elastic moduli prediction of digital rocks}
\author[1]{Hanfeng Zhai\thanks{E-mail: \tt hzhai@stanford.edu}}
\author[1]{Rasool Ahmad}
\author[2]{Tapan Mukerji\thanks{E-mail: \tt mukerji@stanford.edu}}
\author[1]{Wei Cai\thanks{E-mail: \tt caiwei@stanford.edu}}
\affil[1]{\small Department of Mechanical Engineering, Stanford University}
\affil[2]{\small Energy Science and Engineering, Stanford University \vspace{10pt}}

\date{\today}

\begin{document}
\maketitle



\begin{abstract}
Digital rock physics (DRP) aims to estimate effective rock properties (e.g. elastic moduli) directly from 3D micro-CT images. However, direct numerical simulations (DNS) on high-resolution large 3D scans are often computationally prohibitive and severely limits the aplication of DRP.
To address this bottleneck, we combine a lightweight 3D convolutional neural network (CNN) with hierarchical homogenization (HHM) and apply it to determine effective elastic moduli. In this scheme, a large rock image is divided into subcubes. The CNN replaces costly DNS by directly predicting subcube elastic moduli, while HHM upscales subcube-level predictions to the full rock.
Using a shared convolutional backbone, we systematically compare three training targets: (i) full anisotropic $6\times6$ stiffness tensors, (ii) isotropic bulk and shear moduli $(K,G)$, and (iii) Hashin--Shtrikman (HS)-normalized factors.
Across multiple rock types, all three models agree well with DNS results while substantially reducing the computational cost. Moreover, training from scratch on each rock type is fast enough that transfer learning is unnecessary. Across all three targets the accuracy is comparable. In our comparative study, the HS-normalized factor offers the best overall speed--accuracy trade-off while guaranteeing physical consistency, making it a convenient default. The isotropic $(K,G)$ target is a slightly more accurate alternative.


\vspace{10pt}

\end{abstract}

\clearpage
\section{Introduction}

One of the central goals of digital rock physics (DRP) is to predict rock mechanical properties, such as elastic moduli, directly from scans of rock samples \cite{Andr2013_drp_1, Andr2013_drp_2}.
This approach has broad applications in petroleum reservoir characterization, $\text{CO}_2$ sequestration, and mineral resource evaluation.
DRP typically integrates high-resolution microtomographic imaging with numerical simulations to extract effective material properties from voxel-based representations, providing a non-invasive complement to laboratory measurements and enabling detailed analysis of complex pore-scale geometries~\cite{Schepp2020Digital,Lin2019Validation,nishank_mech_estimation}.
Predicting elastic moduli from digital scans involves computing the bulk modulus $K$ and shear modulus $G$ for rocks that are assumed to be isotropic by applying virtual mechanical loads to segmented CT images using quasistatic finite-difference, finite-element method (FEM), or fast Fourier transform (FFT) solvers~\cite{nishank_mech_estimation,Andr2013_drp_1,Andr2013_drp_2}.
%

%
However, balancing the computational cost of simulating large volumes of digital rocks and the need for high spatial resolution remains a key challenge. 
Increasing resolution improves geometric fidelity but makes full-resolution direct numerical simulations (DNS) prohibitively expensive for large rock domains.
Simulating smaller subcubes reduces computational cost but may fail to capture the bulk mechanical response due to heterogeneity and scale effects.
Hierarchical homogenization (HHM) was proposed to mitigate this trade-off by solving elasticity on subcubes and then performing a fast global solve on an assembled coarse representation \cite{Ahmad2022,ahmad2023homogenizing}.
Nevertheless, the initial step in HHM still requires costly DNS on a large number of subcubes, motivating a faster surrogate for this step.
%
%
In parallel, many studies have applied convolutional neural networks (CNNs) to predict rock properties directly from digital scans. The most prominent application is permeability prediction in sandstone and carbonate rocks \cite{Khan2024_cnn_permeability,Tang2022_cnn_permeability,Elmorsy2022_CNN_permeability}.
CNNs have also been used to estimate acoustic properties from CT images \cite{deFigueiredo2019_cnn_p_velo, Karimpouli2019_cnn_s_velo}.
In addition, they have been widely adopted for image processing tasks, including denoising, super-resolution, and reconstruction \cite{Zheng2022_cnn_reconstruction,Bizhani2022_cnn_image_reconstruction}.
For elastic moduli estimation, recent CNN approaches \cite{Hou2022_moduli_cnn, Zorkaltsev2025_moduli_cnn} typically focus on direct prediction at the subvolume scale, but do not address how to scale predictions to much larger rocks.
Ahmad et al.~\cite{ahmad2023homogenizing} took an initial step toward this scaling challenge by using a CNN to replace the DNS step within HHM. Their approach makes two specific modeling choices: (i) treating each subcube as isotropic during assembly and (ii) predicting a Hashin--Shtrikman (HS) bound factor rather than the moduli directly~\cite{Ahmad2022}.
These modeling choices have not been systematically evaluated. The HS-normalized target presumes that each subcube can be treated as isotropic~\cite{Hashin1963}, yet individual subcubes may appear anisotropic even when the full rock is macroscopically isotropic. It is therefore not obvious whether predicting an HS-normalized factor is advantageous in practice, or whether one should instead predict the moduli, or the full stiffness tensor, directly. This motivates a systematic comparison of CNN training targets and their underlying assumptions.

In this work, we assess these modeling choices within the CNN--HHM workflow by comparing CNNs trained on subcubes to predict (i) the full $6\times6$ stiffness tensor (enabling anisotropic subcube behavior), (ii) elastic moduli $(K,G)$ under an isotropic subcube approximation, and (iii) normalized factors constrained by Hashin--Shtrikman (HS) bounds.
{In the following, we refer to them as Model~1, Model~2, and Model~3, respectively}.
We employ early stopping to compare training duration and prediction accuracy across targets.
We also examine whether pretraining and transfer learning are beneficial by comparing against models trained \emph{ad hoc} on each rock type.
Different from previous work~\cite{ahmad2023homogenizing}, our CNN architecture and training procedure are much more efficient and, at the same time, more accurate.
%
We find that directly predicting the elastic moduli under the isotropic assumption (Model~2) improves the accuracy of the predicted overall homogenized moduli. This formulation also makes the model much faster to train.
%
%
We also find that \emph{ad hoc} training on each rock is sufficiently fast that transfer learning does not provide an advantage in practice.

The paper is organized as follows.
In Section \ref{sec:methods}, we describe how HHM is carried out on large digital rock samples, outline our CNN architecture, and detail the training procedure.
In Section \ref{sec:results_discussion}, we present and discuss our results, including evaluation of different CNN training methods on B1 rocks (i.e., Berea sandstone samples as categorized in \cite{Ahmad2022}), prediction performance on other rock types, and practical guidance on model selection for homogenization of elastic moduli in large rocks.
In Section \ref{sec:conclusion}, we offer some concluding remarks.

\section{Methods\label{sec:methods}}

\begin{figure}[htbp]
    \centering
    \includegraphics[width=.3\linewidth]{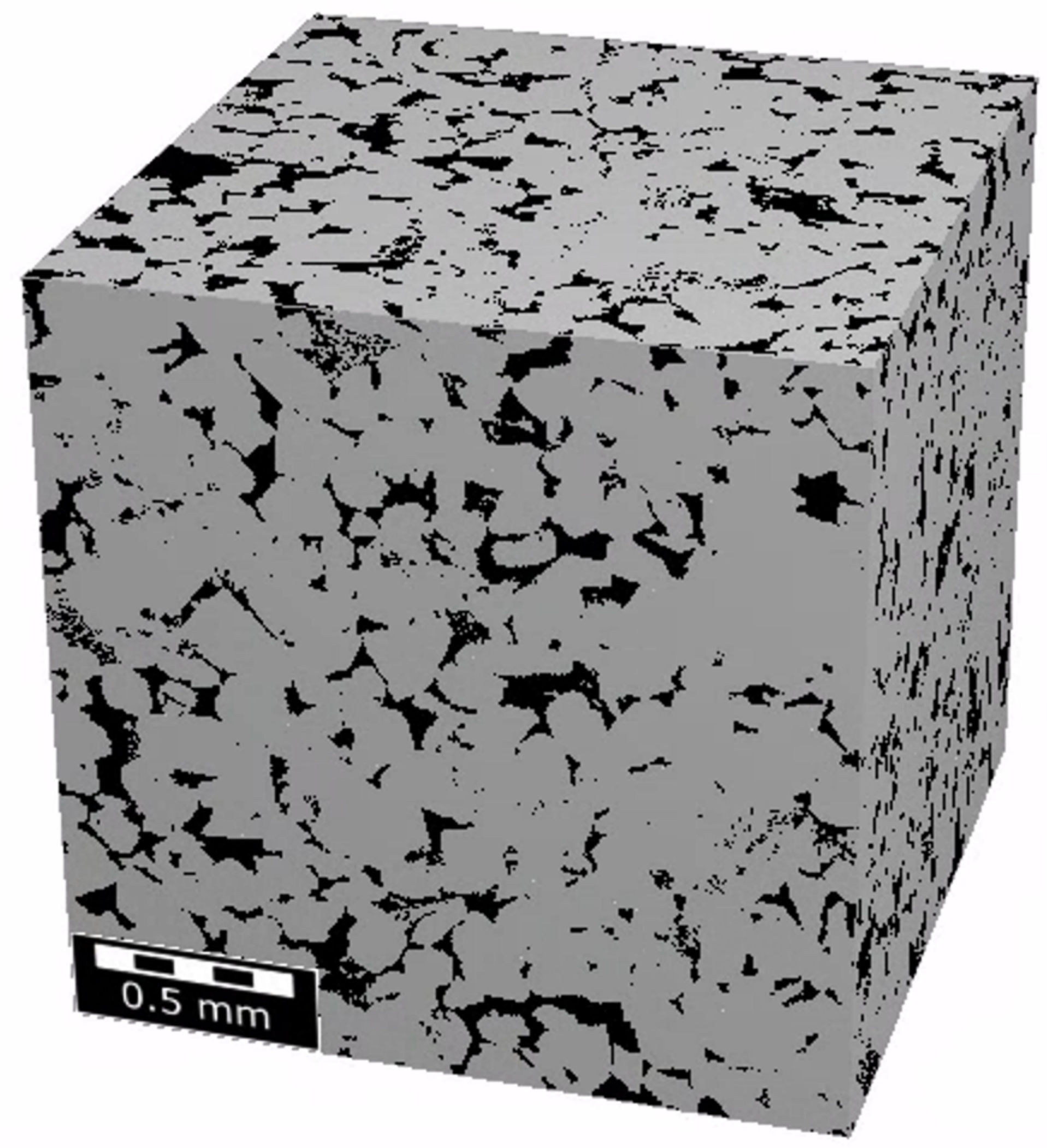}
    \caption{Representative 3D rendering of the B1 digital rock at $900\times900\times900$ voxels. Solid voxels (gray) denote the mineral frame and void voxels (black) denote pores; this sample is used throughout the paper to illustrate the CNN--HHM workflow and large-volume prediction behavior.}
    \label{fig:b1}
\end{figure}

{We analyze high-resolution micro-CT images from five digital rock types: B1 and B2 (Berea sandstones), CG (Castlegate sandstone),
and FB1 and FB2 (Fontainebleau sandstones), following the dataset in prior work \cite{ahmad2023homogenizing}.}
%
%
Figure~\ref{fig:b1} shows a representative 3D rendering of the B1 sample used as the primary demonstration case in this work.

{For reference, reported sample porosities include} $\phi_{\mathrm{B1}}=16.51\%$ and $\phi_{\mathrm{CG}}=22.20\%$ \cite{Ahmad2022}; the other samples (B2, FB1, FB2) are taken from the same curated sandstone dataset and are analyzed with the same segmentation and homogenization pipeline.
Each specimen is represented as a segmented binary voxel volume of size $900\times900\times900$, where each voxel belongs to either mineral or pore phase.

\begin{figure}[htbp]
    \centering
    \includegraphics[width=\linewidth]{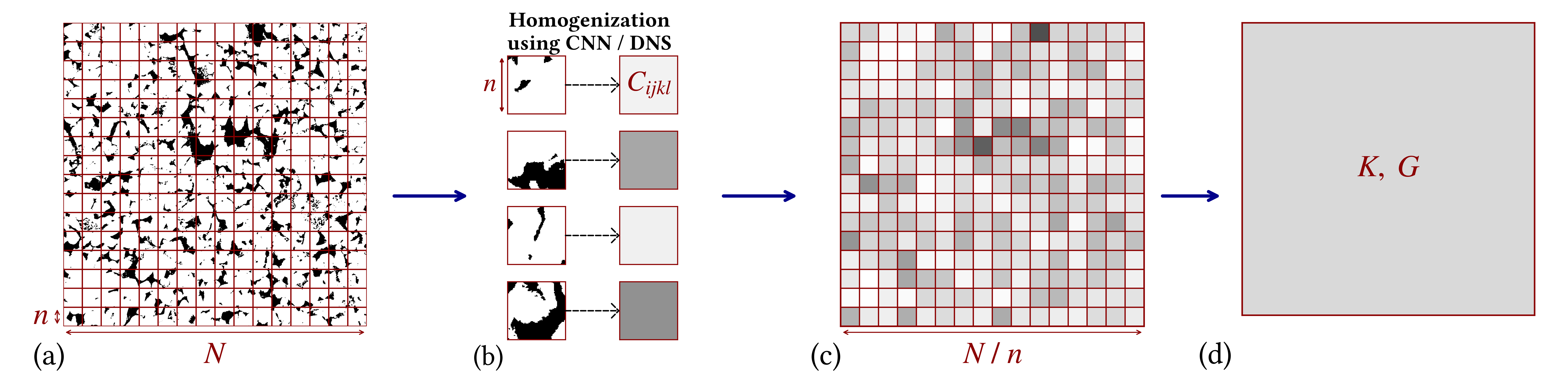}
    \caption{CNN--HHM workflow: (a) partition the 3D scan into subcubes, (b) compute subcube effective elasticity (DNS or CNN surrogate), (c) assemble a coarse-grained elastic volume, and (d) solve global elasticity on the coarse domain to obtain effective bulk and shear moduli.}
    \label{fig:hhm_cnn_schematic}
\end{figure}

To predict the homogenized elastic moduli for much larger rock cubes, we use hierarchical homogenization (HHM), as summarized in Figure~\ref{fig:hhm_cnn_schematic}.
For an $N\times N\times N$ volume, HHM partitions the rock into $(N/n)^3$ non-overlapping $n\times n\times n$ subcubes, computes the effective elastic properties for each subcube, assembles these local responses into a coarse rock, and solves one global elasticity problem on the assembled domain.
Subcube size must exceed the rock correlation length so that local homogenization remains statistically representative. 
Using spatial correlation analysis from Ref.~\cite{ahmad2023homogenizing}, we set $n=75$ and use $75\times75\times75$ subcubes throughout.
{For large-volume validation, we crop $300\times300\times300$ and $600\times600\times600$ subregions from each $900\times900\times900$ parent rock and apply CNN--HHM to these crops, comparing predictions against DNS references computed on the same cropped regions.}

\begin{table}[htbp]
    \centering
    \caption{Phase elastic parameters used in the FFT-based subcube homogenization step: $E$ is the Young's modulus and $\nu$ is the Poisson's ratio. The mineral matrix is modeled as a solid phase, and the pore phase (air) is approximated as a very compliant solid. {The solid-phase properties are chosen to resemble quartz, consistent with the sandstone nature of all analyzed samples}.}
    \begin{tabular}{l|c c}
    \toprule
     & Mineral & Air (Pore)\\ \midrule
        $E$\quad[GPa] & 95.29 & $2.90\times10^{-4}$ \\
        $\nu$\quad[-] & 0.05 & 0.45 \\ \bottomrule
    \end{tabular}
    \label{tab:elastic_param_calculation}
\end{table}

We compute each subcube elastic response with an FFT-based solver, {using GeoDict \cite{GeoDict_software} for step~1 of HHM, with fixed mineral and pore elastic properties} listed in Table~\ref{tab:elastic_param_calculation} \cite{Ahmad2022, Moulinec1998_fft_solver, GeoDict_software}. Here, air is approximated as a very compliant solid when computing homogenized moduli, which is a good-enough approximation for our rocks and scale of interest.
No padding is used in this subcube-scale elasticity solve.

In our formulation, {there are 3 different models, named Model 1, Model 2, and Model 3, based on different output of the CNN models}. 
%
For every subcube, DNS outputs a full $6\times6$ anisotropic stiffness matrix $\bm{\mathcal{C}}$, and these stiffness matrices are used as targets for Model~1. The local constitutive relation is written in Voigt form as
\begin{equation}
\bm{\sigma}=\bm{\mathcal{C}}\,\bm{\varepsilon},
\label{eq:constitutive_voigt_methods}
\end{equation}
where $\bm{\sigma}$ and $\bm{\varepsilon}$ are the volume-averaged stress and strain vectors for each subcube.

While the rocks are macroscopically isotropic, we would like to assess whether this isotropic approximation is acceptable at the subcube scale. 
When isotropy is acceptable, from the resulting stiffness tensor $\bm{\mathcal{C}}$, one can derive $K$ and $G$ \cite{Goldfarb2022Predictive,Lin2019Validation}, using different average schemes such as Voigt~\cite{voigt1928kristallphysik}, Reuss~\cite{reuss1929}, or Hill~\cite{hill1952} based on linear elastic and isotropic assumptions.
In this work, each $\bm{\mathcal{C}}$ is reduced to Voigt-averaged bulk and shear moduli $(K,G)$, used as Model~2 targets, with
\begin{equation}
\begin{aligned}
K_\mathrm{V} &= \frac{1}{9}\big(C_{11}+C_{22}+C_{33}+2(C_{12}+C_{13}+C_{23})\big),\\
G_\mathrm{V} &= \frac{1}{15}\big(C_{11}+C_{22}+C_{33}-C_{12}-C_{13}-C_{23}+3(C_{44}+C_{55}+C_{66})\big).
\end{aligned}
\label{eq:voigt_KG_methods}
\end{equation}

If we assume each subcube is elastically isotropic, Hashin--Shtrikman (HS) bounds~\cite{hashin1962some} define porosity-dependent upper and lower limits for effective bulk and shear moduli in a two-phase solid--pore composite. Let porosity be $\phi$, with phase moduli $(K_m,G_m)$ for mineral and $(K_p,G_p)$ for pore, then
%
\begin{equation}
\begin{aligned}
K^{\mathrm{HS}}_{\mathrm{upper}} &= K_m + \frac{1-\phi}{(K_p-K_m)^{-1}+\phi\,(K_m+\tfrac{4}{3}G_m)^{-1}},\\
K^{\mathrm{HS}}_{\mathrm{lower}} &= K_p + \frac{\phi}{(K_m-K_p)^{-1}+(1-\phi)\,(K_p+\tfrac{4}{3}G_p)^{-1}},\\
G^{\mathrm{HS}}_{\mathrm{upper}} &= G_m + \frac{1-\phi}{(G_p-G_m)^{-1}+2\phi\,(K_m+2G_m)\,[5G_m(K_m+\tfrac{4}{3}G_m)]^{-1}},\\
G^{\mathrm{HS}}_{\mathrm{lower}} &= G_p + \frac{\phi}{(G_m-G_p)^{-1}+2(1-\phi)\,(K_p+2G_p)\,[5G_p(K_p+\tfrac{4}{3}G_p)]^{-1}}.
\end{aligned}
\label{eq:hs_bounds_methods}
\end{equation}
Predictions outside these bounds are unphysical, so HS consistency provides a physics-based reliability constraint.

For Model~3, we normalize effective moduli into factors in $[0,1]$ using the HS bounds:
\begin{equation}
\begin{aligned}
f_K &= \frac{K^{\mathrm{HS}}_{\mathrm{upper}}-K_{\mathrm{V}}}{K^{\mathrm{HS}}_{\mathrm{upper}}-K^{\mathrm{HS}}_{\mathrm{lower}}},\\
f_G &= \frac{G^{\mathrm{HS}}_{\mathrm{upper}}-G_{\mathrm{V}}}{G^{\mathrm{HS}}_{\mathrm{upper}}-G^{\mathrm{HS}}_{\mathrm{lower}}}.
\end{aligned}
\label{eq:hs_normalized_factors}
\end{equation}
By construction, $f_K,f_G\in[0,1]$ if and only if predictions satisfy HS bounds, and $(f_K,f_G)$ are used as Model~3 targets.

\begin{figure}[htbp]
    \centering
    \includegraphics[width=.7\linewidth]{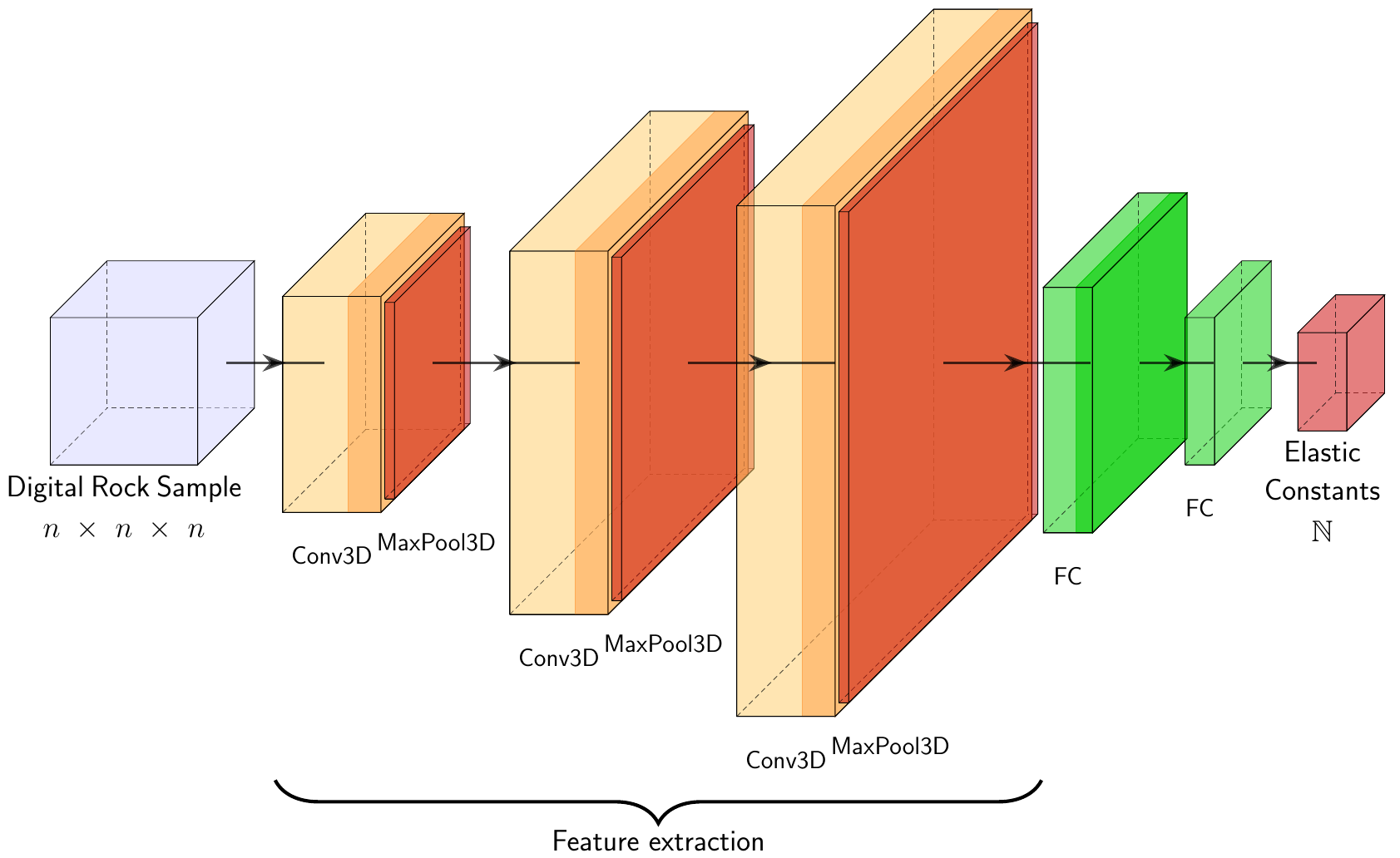}
    \caption{3D CNN surrogate used to predict subcube elasticity from voxelized micro-CT inputs. Convolution + max-pooling blocks extract multiscale features, followed by fully connected layers that output either the full $6\times6$ stiffness tensor, $(K,G)$, or HS-normalized factors.
    The feature extraction section is shared among Models 1, 2, and 3.
    See Table~\ref{tab:cnn3d_arch} for the size of the neural network layers.
    }
    \label{fig:cnn_schematic}
\end{figure}

All three models share the same lightweight 3D-CNN backbone, where ``backbone'' denotes the common feature-extraction section reused across different output targets; this architecture is illustrated in Figure~\ref{fig:cnn_schematic}.
Specifically, three Conv3D$+$MaxPool blocks progressively extract features, followed by a shared fully connected representation and a task-specific linear output head of size $\mathbb{N}$.
{The final dark-red layer in Figure~}\ref{fig:cnn_schematic} { represents these three alternative output choices. Model~1 predicts the stiffness tensor components ($\mathbb{N}=36$, or $21$ independent components). Model~2 predicts the effective bulk and shear moduli $(K,G)$ directly ($\mathbb{N}=2$). Model~3 predicts the HS-normalized factors $(f_K,f_G)$ that are subsequently mapped back to $(K,G)$ ($\mathbb{N}=2$).}
Using the layer dimensions in Table~\ref{tab:cnn3d_arch}, the feature-extraction section (three Conv3D blocks) has only $1.
18\times10^5$ trainable parameters, and the total trainable parameters are $5.68\times10^7$ (Model~1) and $5.67\times10^7$ (Models~2, 3).
{We therefore use ``lightweight'' in a practical architecture sense (shallow depth, simple operators, and a compact convolutional feature extractor),} especially relative to the heavier BatchNorm/Dropout/PReLU design used by Ahmad et al.~\cite{ahmad2023homogenizing}.

For a fair comparison, we train all three models with mean-squared-error loss and the Adam optimizer under identical settings: batch size $8$, learning rate $0.001$, and maximum $100$ epochs with an 80\%/20\% train-test split. Early stopping is applied using the PyTorch validation-loss plateau criterion, i.e., training stops when validation loss plateaus or increases over recent epochs.
The final reported model is the one that exhibits the lowest test loss during the training period.

\section{Results and discussions\label{sec:results_discussion}}

\begin{figure}[htbp]
    \centering
    \begin{subfigure}[t]{0.355\linewidth}
        \centering
        \includegraphics[width=\linewidth]{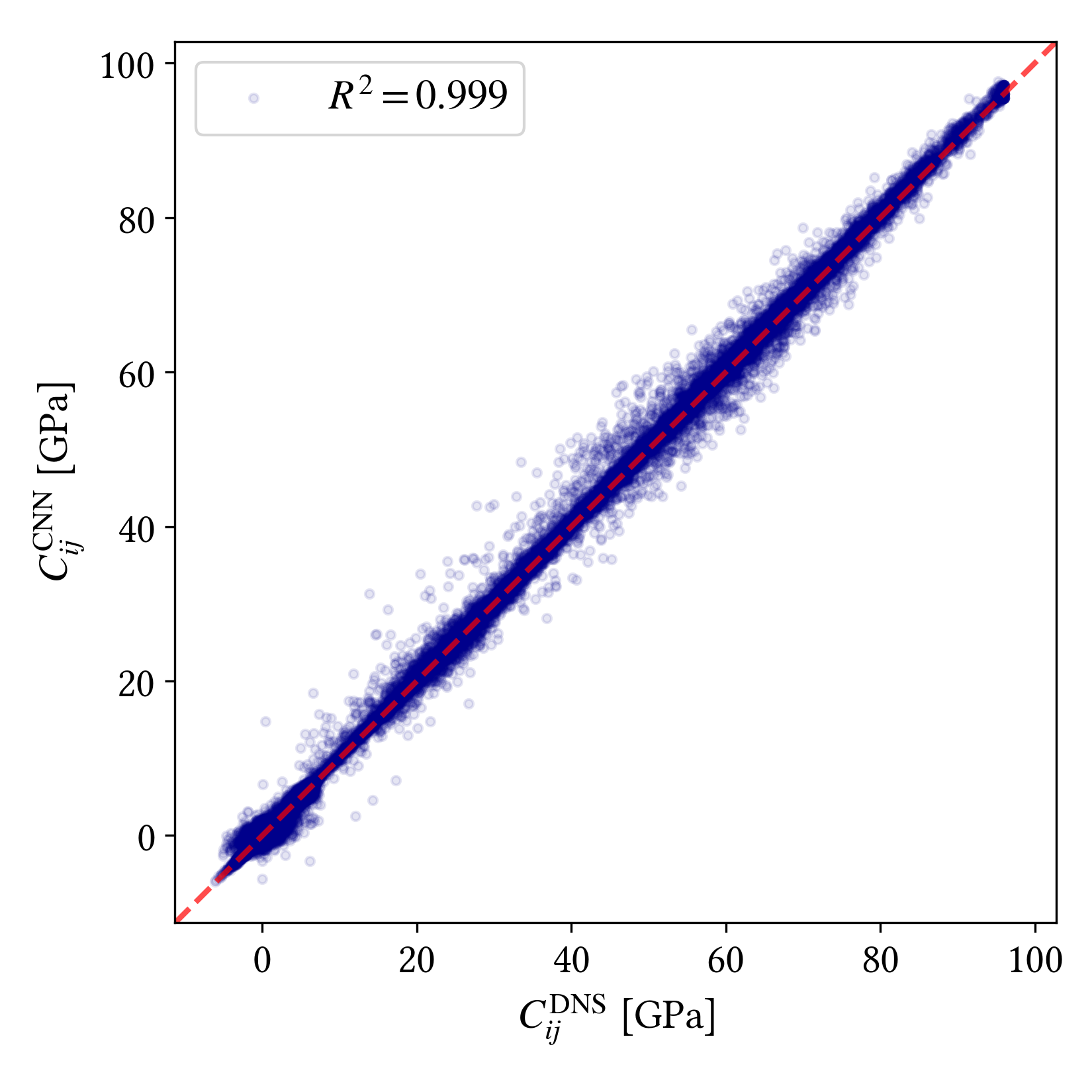}
        \caption{}
    \end{subfigure}
    \hfill
    \begin{subfigure}[t]{0.295\linewidth}
        \centering
        \includegraphics[width=\linewidth]{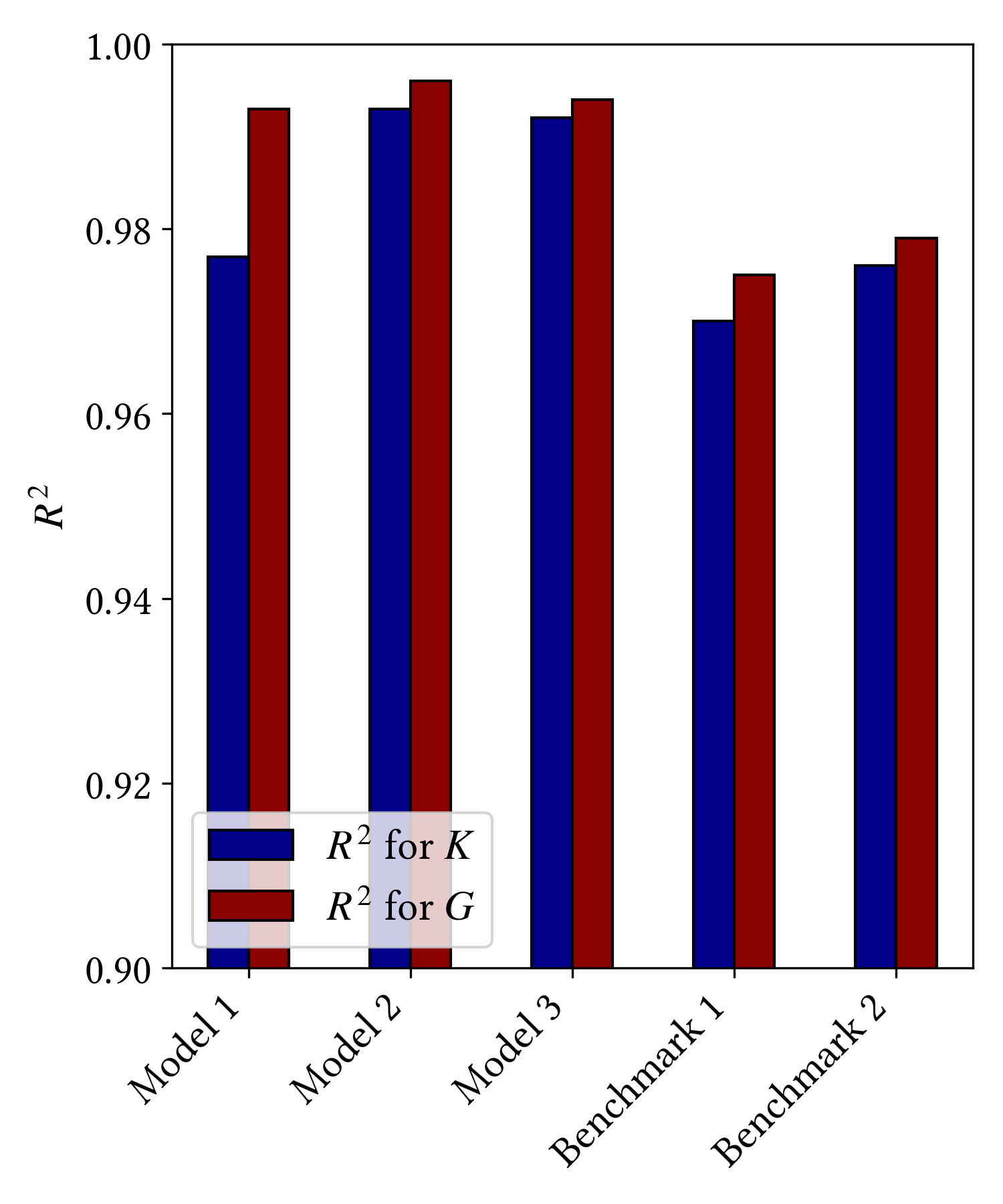}
        \caption{}
    \end{subfigure}
    \hfill
    \begin{subfigure}[t]{0.295\linewidth}
        \centering
        \includegraphics[width=\linewidth]{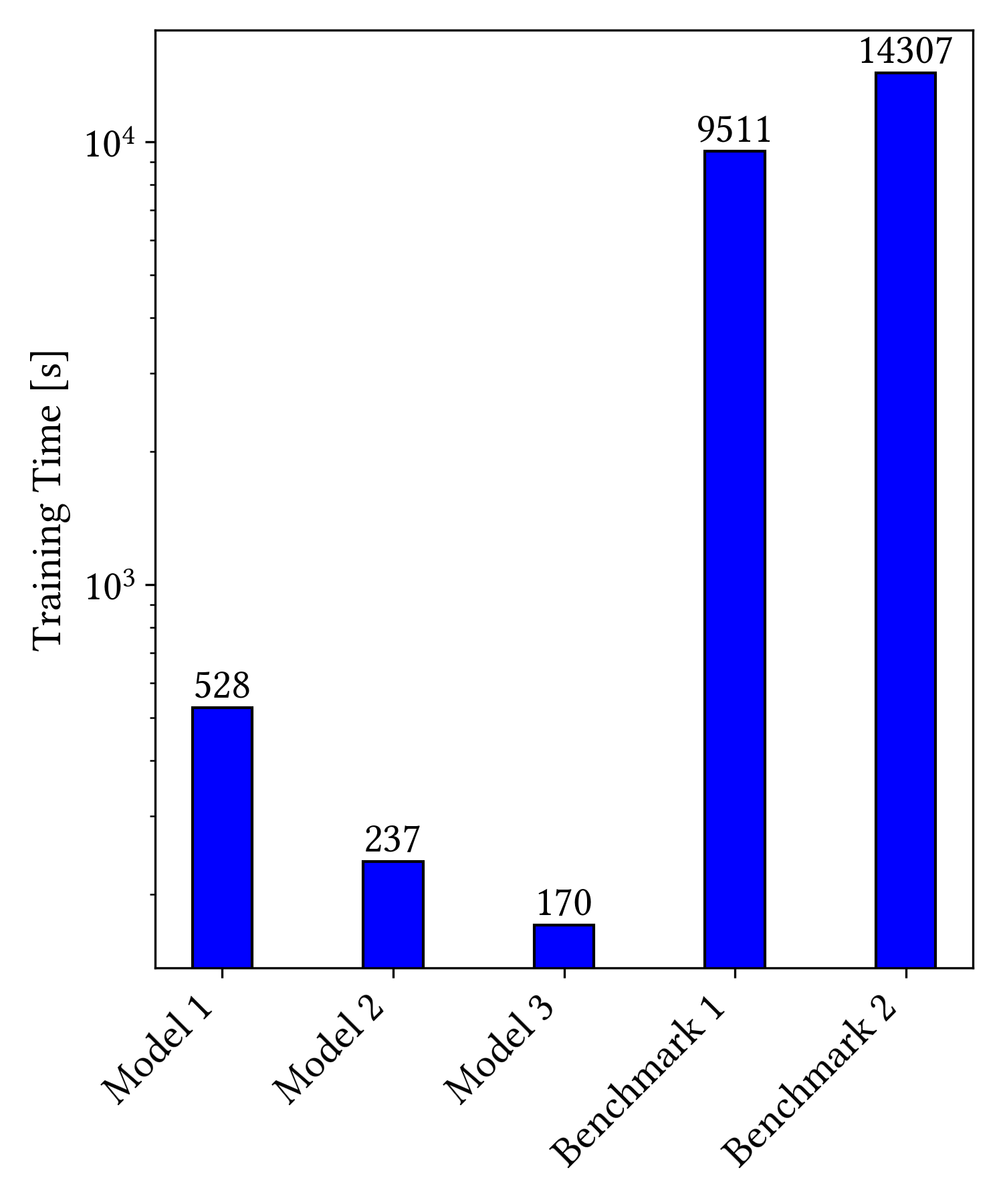}
        \caption{}
    \end{subfigure}
    \caption{
    B1 subcube training evaluation and stiffness-level verification. (a) Parity plot comparing CNN-predicted and DNS-reference values for the 21 independent components of the stiffness tensor (Voigt notation). Each marker corresponds to one tensor component from one test subcube, and the $45^\circ$ diagonal denotes perfect agreement. {(b) Test-set $R^2$ for bulk modulus $K$ and shear modulus $G$ for Model~1, Model~2, Model~3, Benchmark~1, and Benchmark~2 (Ref.~\cite{ahmad2023homogenizing}). (c) Corresponding training time (seconds) for the same five approaches, under identical hardware and early-stopping settings.}}
    \label{fig:eval_B1_cnn_training}
\end{figure}

For rock B1, we first evaluate Model~1 using the parity plot of all 21 independent stiffness constants (Figure~\ref{fig:eval_B1_cnn_training}(a)), where each point compares a CNN-predicted stiffness entry with the DNS reference and the diagonal indicates perfect agreement. We report the mean-squared error (MSE) over these 21 components as the stiffness-level summary metric.
The prediction accuracy is high ($R^2=0.999$), indicating that most stiffness-tensor components are recovered accurately.

\begin{figure}[htbp]
    \centering
    \includegraphics[width=.9\linewidth]{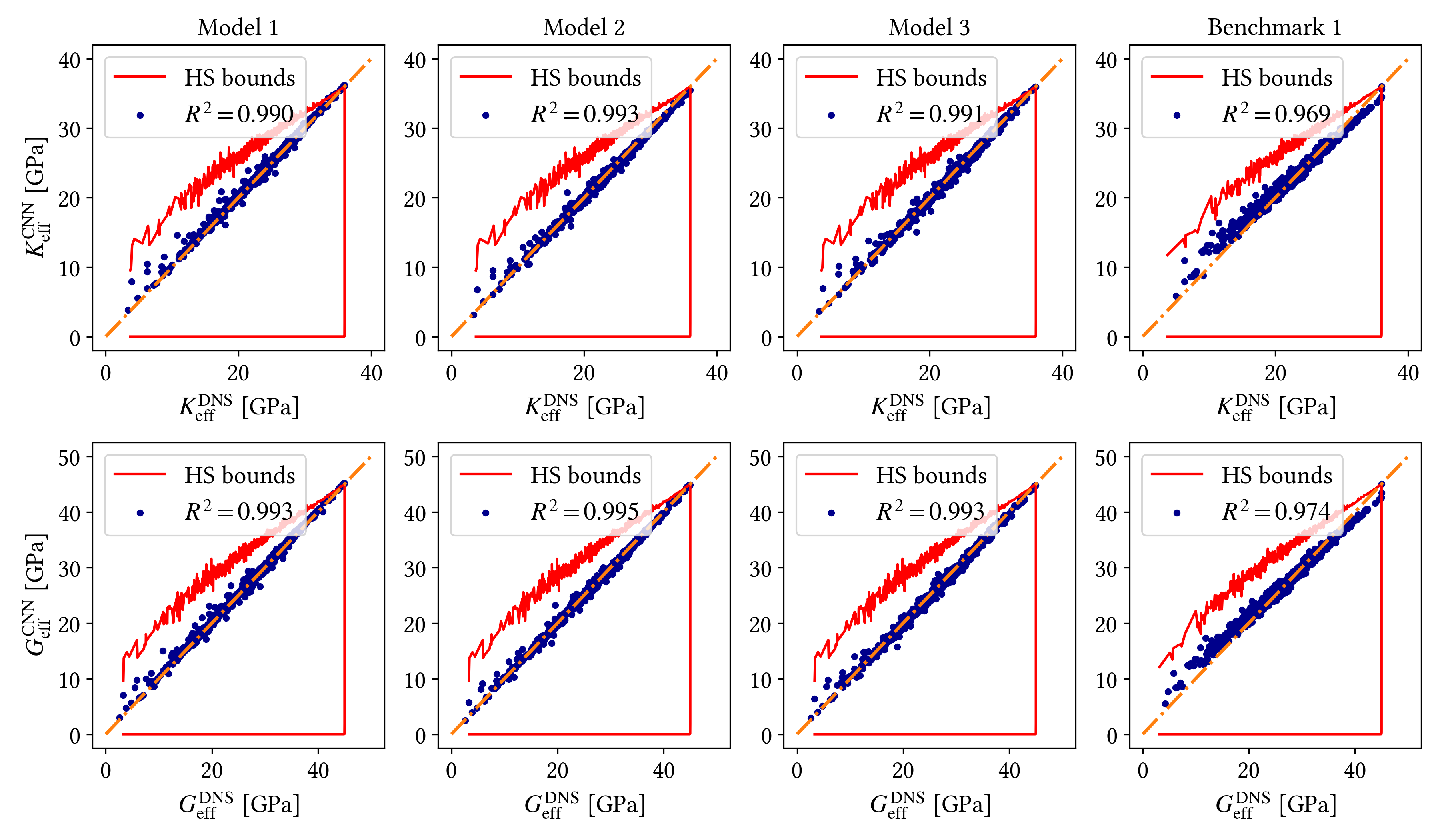}\\
    {\hspace{10pt}\centering(a)\hspace{95pt}(b)\hspace{95pt}(c)\hspace{95pt}(d)}
    \caption{B1 subcube parity plots for bulk modulus $K$ (top row) and shear modulus $G$ (bottom row). {Each column corresponds to (a) Model~1, (b) Model~2, (c) Model~3, and (d) Benchmark~1. Solid red lines mark the Hashin--Shtrikman bounds.}}
    \label{fig:cnn_B1_training_meth}
\end{figure}

For a fair subcube-level comparison of Models~1--3 on common targets, we post-process Model~1 by Voigt-averaging its predicted stiffness tensor to obtain $(K,G)$, then compare these values against the direct $(K,G)$ predictions of Models~2 and~3 using DNS references (Figure~\ref{fig:eval_B1_cnn_training}(b,c); Figure~\ref{fig:cnn_B1_training_meth}).
{For comparison, we additionally include two CNN models as benchmarks based on Ahmad et al.}~\cite{ahmad2023homogenizing} {that predict HS-normalized factors, similar to our Model~3. Benchmark~1 uses their architecture retrained using the same epochs, batch size, and early stopping as Models~1--3 in this work. Benchmark~2 uses the original training setup as reported in Ref.~\cite{ahmad2023homogenizing}. Together, Models~1--3 and these two benchmarks form the five approaches shown in Figure~}\ref{fig:eval_B1_cnn_training}{(b,c).}
Under the same early-stopping protocol, we also compare total training time.
Figure~\ref{fig:eval_B1_cnn_training}(c) shows that predicting HS-normalized factors (Model~3) yields the shortest training time. Figure~\ref{fig:eval_B1_cnn_training}(b) shows that direct $(K,G)$ prediction (Model~2) and HS-normalized prediction (Model~3) achieve the highest accuracy in terms of $R^2$.
{Benchmark~1 and Benchmark~2 both require substantially longer training time than Models~1--3, reflecting the heavier architecture and training protocol of Ref.~}\cite{ahmad2023homogenizing}.

Figure~\ref{fig:cnn_B1_training_meth} further confirms that {Models~1--3 and Benchmark~1 all} produce accurate subcube-level modulus predictions, with most samples lying within the HS bounds. Our three models achieve $R^2>0.99$, whereas {Benchmark~1 is around $R^2\approx0.97$}. 
Our three models do not exhibit the systematic deviation of the slope in the parity plots, which is discernible in Benchmark 1 (Figure~\ref{fig:cnn_B1_training_meth}(d)), i.e., over-prediction for the low-modulus subcubes and under-predictions for the high-modulus subcubes.
Together with Figure~\ref{fig:eval_B1_cnn_training}, these results indicate that both direct $(K,G)$ and HS-normalized targets are effective, while the HS-normalized target provides a modest speed advantage.
From {Figure~\ref{fig:cnn_B1_training_meth}} (a) and (b),
we can see that although HS bounds are not enforced in Models 1 and 2, most predictions still satisfy the HS bounds well.
The faster convergence of the HS-normalized target (Model~3) can be attributed to its bounded output range $[0,1]$, which improves the numerical conditioning of the loss and facilitates gradient-based optimization.
More broadly, this illustrates that injecting physical knowledge into the model (here, by normalizing against the HS bounds) can accelerate training.
We then evaluate the prediction accuracies
for the remaining four rocks (see Figure~\ref{fig:all_rock_slices} and Tables~\ref{tab:r2_300} to~\ref{tab:rmse_600})
to assess consistency and cross-dataset generalization.
All the above observations on rock B1 also hold for other four rocks.
%
%


\begin{figure}[h]
    \centering
    \includegraphics[width=.65\linewidth]{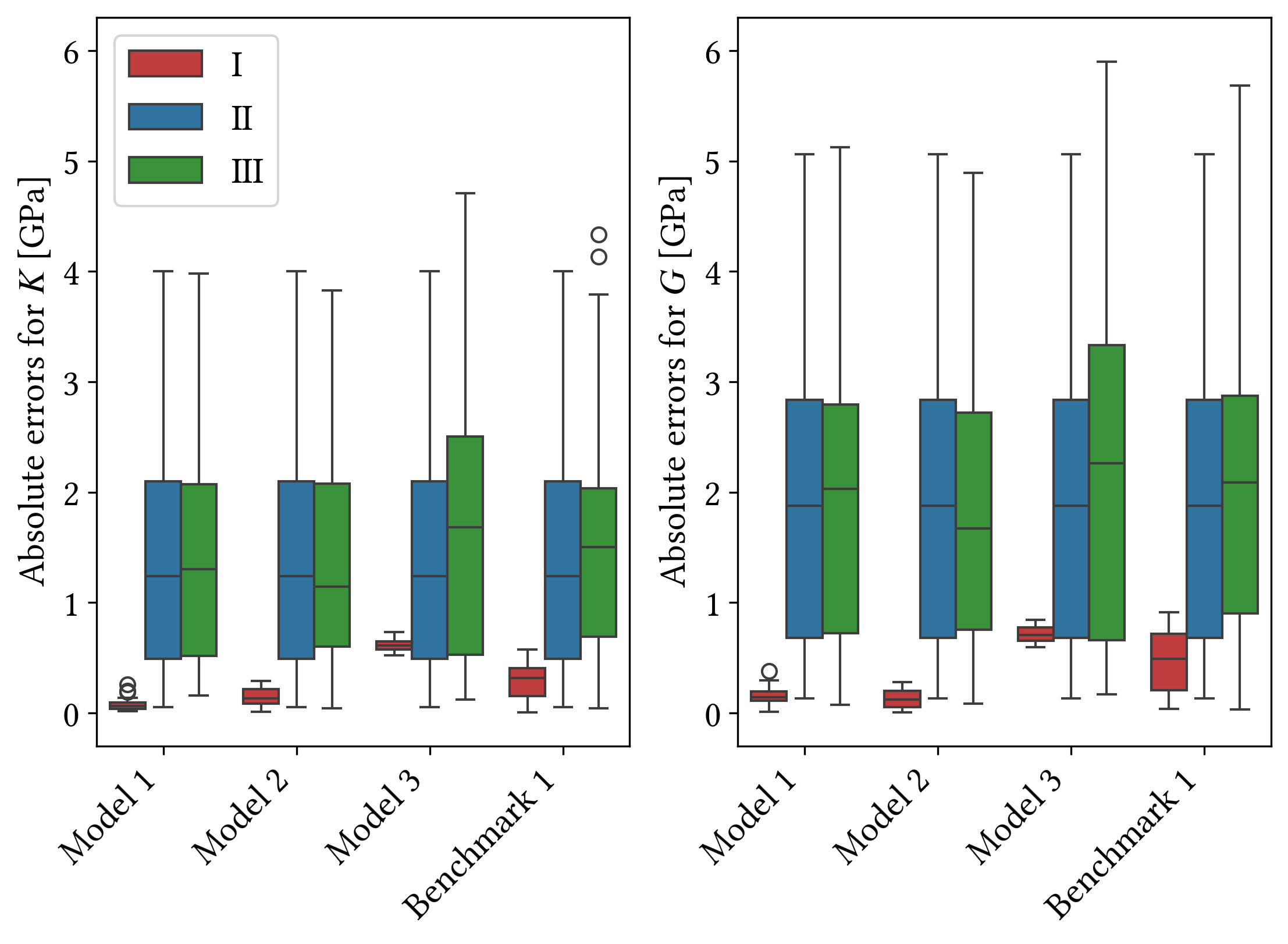}\\
    {\hspace{30pt}\centering(a)\hspace{170pt}(b)}
    \caption{Error propagation in the CNN--HHM pipeline for a {$300\times300\times300$ crop from the $900\times900\times900$ B1 parent rock}. Box plots summarize absolute-error distributions for (a) bulk modulus $K$ and (b) shear modulus $G$. {The $x$-axis groups correspond to Model~1, Model~2, Model~3, and Benchmark~1.} In each panel, groups I, II, and III denote the three comparison scenarios, respectively: I (red) is CNN vs.~DNS at the subcube level (surrogate error), II (blue) is HHM (using DNS) vs.~DNS at the large-volume level (upscaling error), and III (green) is HHM (using CNN) vs.~DNS at the large-volume level (total pipeline error).}
    \label{fig:eval_errors_CNN_HHM}
\end{figure}

Figure~\ref{fig:eval_errors_CNN_HHM} shows the different error contributions in the CNN--HHM workflow.
For both bulk and shear moduli, the CNN vs.~DNS subcube error remains consistently smaller than the HHM vs.~DNS large-volume discrepancy. In other words, the error from replacing DNS with the CNN surrogate at the subcube level (including the isotropic subcube approximation) is much smaller than the unavoidable error introduced by HHM upscaling itself, so the CNN contributes little to the total error in the predicted large-volume moduli.

\begin{table}[htbp]
\centering
\caption{{Mean absolute errors for a $300\times300\times300$ crop from the $900\times900\times900$ B1 parent rock, comparing Models~1--3 and Benchmark~1.} Values are reported in the format of $K/G$ in GPa.}
\begin{tabular}{c|cccc}
\toprule
 & Model 1 & Model 2 & Model 3 & {Benchmark 1} \\
\midrule
CNN vs.~HHM
 & 0.080 / 0.158
 & 0.144 / 0.124
 & 0.614 / 0.710
 & 0.292 / 0.470 \\

CNN vs.~DNS
 & 1.513 / 2.036
 & 1.453 / 1.960
 & 1.748 / 2.315
 & 1.575 / 2.167 \\

HHM vs.~DNS
 & 1.516 / 2.026
 & 1.516 / 2.026
 & 1.516 / 2.026
 & 1.516 / 2.026 \\

\bottomrule
\end{tabular}
\label{tab:error_analysis}
\end{table}

Table~\ref{tab:error_analysis} quantifies the three comparisons shown in Figure~\ref{fig:eval_errors_CNN_HHM}: CNN vs.~HHM, CNN vs.~DNS, and HHM vs.~DNS. Consistent across all three models, the additional error associated with the CNN surrogate and isotropic subcube approximation is smaller than the discrepancy between HHM and DNS.
{Interestingly, Model~3 sometimes achieves slightly lower error than Model~2, which is not expected from a strict error-propagation standpoint.} Because the surrogate and upscaling errors need not accumulate, they can partially cancel, so CNN--HHM can match the accuracy of raw HHM at a fraction of the runtime.

\begin{figure}[h]
    \centering
    \includegraphics[width=.9\linewidth]{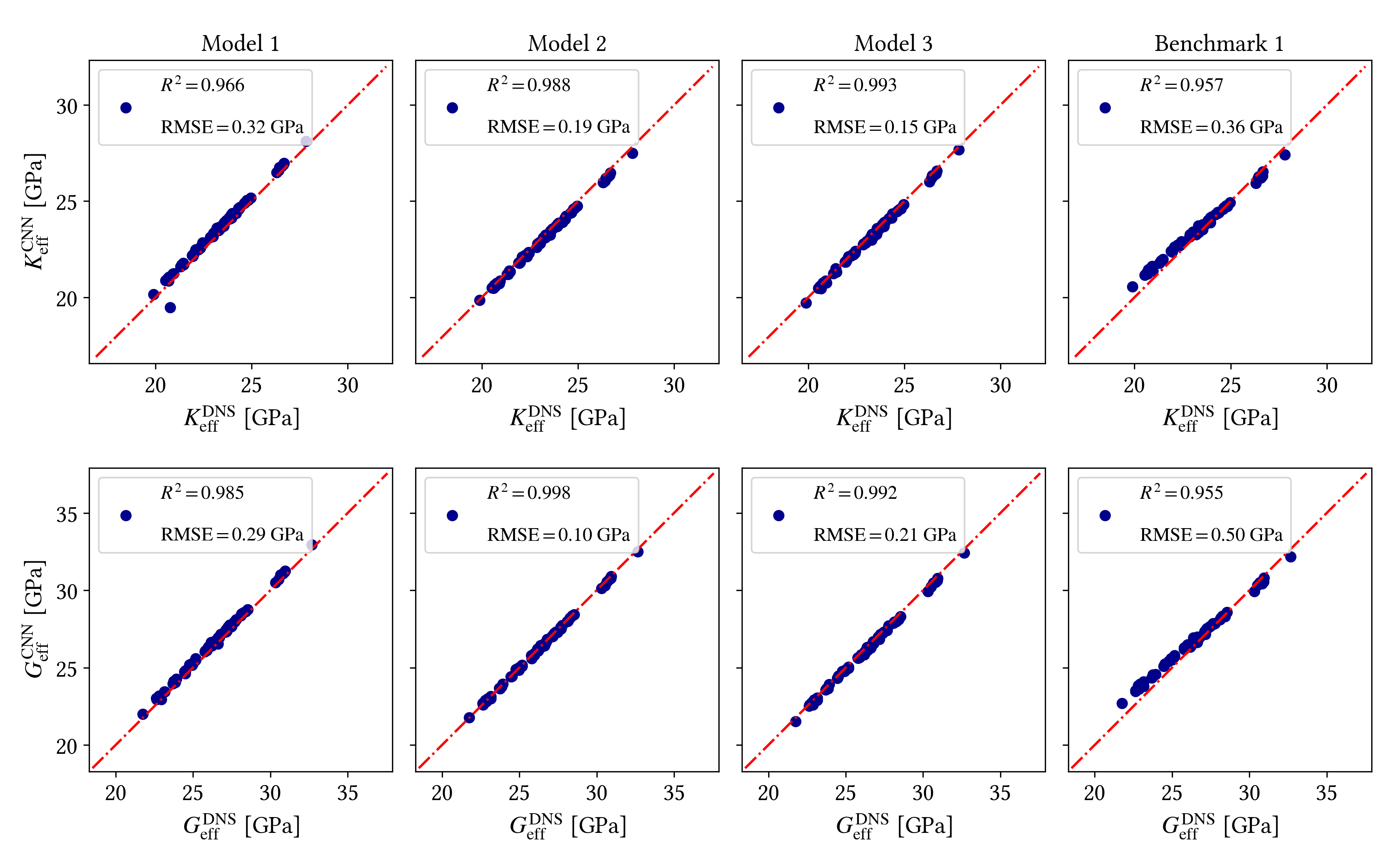}\\
    \caption{{CNN--HHM prediction on a $300\times300\times300$ crop taken from the $900\times900\times900$ B1 parent rock.} Parity plots compare predicted effective bulk modulus $K$ and shear modulus $G$ against DNS references for Model~1, Model~2, Model~3, and Benchmark~1.}
    \label{fig:cnn_B1_hhm_300}
\end{figure}

Figure~\ref{fig:cnn_B1_hhm_300} compares the CNN--HHM predicted bulk and shear moduli against DNS references for the {$300\times300\times300$ crop from the B1 $900\times900\times900$ rock}, and the agreement is high. Models~2 and~3 (direct $(K,G)$ and HS-normalized targets) both achieve $R^2>0.99$, while Model~1 has a slightly lower $R^2$.
This trend suggests that, for this macroscopically isotropic rock, isotropic-target training of the subcubes (Models~2 and~3) is advantageous within CNN--HHM.
All three of our models outperform {Benchmark~1}.

\begin{figure}[h]
    \centering
    \includegraphics[width=.9\linewidth]{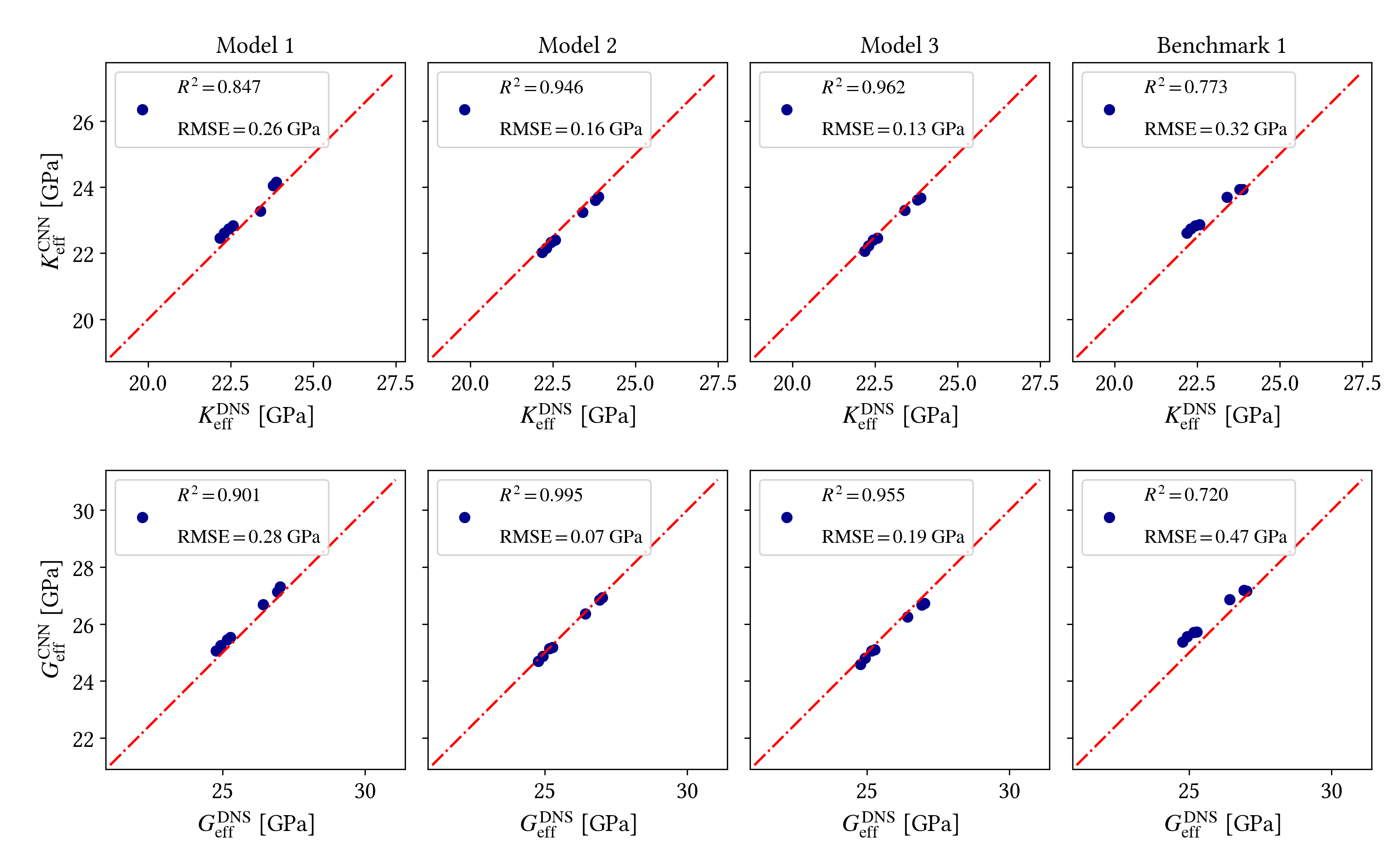}\\
    \caption{{CNN--HHM prediction on a $600\times600\times600$ crop taken from the same $900\times900\times900$ B1 parent rock. Parity plots use the same Model~1--3 and Benchmark~1 labeling as in Figure~}\ref{fig:cnn_B1_hhm_300}.}
    \label{fig:cnn_B1_hhm_600}
\end{figure}

Figure~\ref{fig:cnn_B1_hhm_600} shows that the same CNN--HHM workflow remains accurate when applied to the {$600\times600\times600$ crop from the same $900\times900\times900$ parent volume}. Models~2 and~3 again provide the strongest performance, whereas {Benchmark~1} yields noticeably lower accuracy.
Similar HHM-level trends are observed for the other rocks, where CNN--HHM predictions for both bulk and shear moduli remain close to DNS for B2, CG, FB1, and FB2, supporting robust transfer of the workflow across different rock types (See Tables~\ref{tab:r2_300}--\ref{tab:rmse_600}).
The results in Figures~\ref{fig:cnn_B1_hhm_300} and~\ref{fig:cnn_B1_hhm_600} suggest that combining the isotropic assumption (Models~2 and~3) with HHM can predict overall moduli of larger rocks more accurately, as these approaches systematically outperform alternatives across both {$300\times300\times300$ and $600\times600\times600$ crops from the $900\times900\times900$ parent rock}. This also suggests that assuming isotropic subcubes is a reasonable approach when evaluating much larger rocks.

{We next report CNN--HHM runtimes, decomposing the wall-clock time as}
\(t_{\mathrm{tot}}=t_{\mathrm{CNN}}+t_{\mathrm{prep}}+t_{\mathrm{HHM}}\), where
\(t_{\mathrm{CNN}}\) is surrogate inference, \(t_{\mathrm{prep}}\) is data preparation/conversion,
and \(t_{\mathrm{HHM}}\) is the coarse-scale homogenization solve.
Table~\ref{tab:time_b1_300_cpu_gpu} reports this breakdown for the $300\times300\times300$ crop of the B1 rock under both CPU- and GPU-based CNN inference.

\begin{table}[htbp]
    \centering
    \caption{Runtime breakdown on B1 ($300\times300\times300$) for the CNN--HHM pipeline. All times are in seconds.}
    \label{tab:time_b1_300_cpu_gpu}
    \small
    \setlength{\tabcolsep}{4pt}
    \begin{tabular}{lcccccc}
    \hline
    & \multicolumn{2}{c}{$t_{\mathrm{CNN}}$}
    & \multicolumn{2}{c}{$t_{\mathrm{HHM}}$}
    & \multicolumn{2}{c}{$t_{\mathrm{tot}}$} \\
    \cline{2-7}
    Approach & CPU & GPU & CPU & GPU & CPU & GPU \\
    \hline
    Model 1 
        & $7.73\times10^{-1}$ & $2.75\times10^{0}$
        & $1.19\times10^{-1}$ & $7.67\times10^{-2}$
        & $9.14\times10^{-1}$ & $2.85\times10^{0}$ \\

    Model 2 
        & $1.24\times10^{0}$ & $1.35\times10^{-1}$
        & $1.19\times10^{-1}$ & $7.64\times10^{-2}$
        & $1.38\times10^{0}$ & $2.32\times10^{-1}$ \\

    Model 3 
        & $1.05\times10^{0}$ & $2.95\times10^{-1}$
        & $1.27\times10^{-1}$ & $8.20\times10^{-2}$
        & $1.20\times10^{0}$ & $3.98\times10^{-1}$ \\

    Benchmark 1~\cite{ahmad2023homogenizing}
        & $2.19\times10^{1}$ & $1.12\times10^{0}$
        & $1.11\times10^{-1}$ & $7.69\times10^{-2}$
        & $2.21\times10^{1}$ & $1.22\times10^{0}$ \\
    \hline
    \end{tabular}
\end{table}



{Across hardware settings, our three models are consistently faster than the baseline in total runtime.}
With CPU inference, our pipeline reduces total time by about one order of magnitude relative to Ref.~\cite{ahmad2023homogenizing}; with GPU inference, Model~2 and Model~3 remain faster while preserving high prediction accuracy. Compared with the reported direct full-rock DNS cost on the order of $10^5$ seconds \cite{Ahmad2022}, the CNN--HHM workflow provides a significant acceleration for large-rock evaluation, by a factor on the order of $10^5$--$10^6$.


In practice, for macroscopically isotropic rocks, {we would recommend Model~3 across all five rock types evaluated here because it provides the best computational efficiency while maintaining high predictive quality at both the subcube level and within the HHM upscaling workflow (Tables~}\ref{tab:r2_300}{--}\ref{tab:rmse_600}). Model~2 is also a strong alternative, especially when direct $(K,G)$ outputs are preferred. Model~3's HS-normalized targets enforce physical consistency and, in our experiments, deliver a strong speed--accuracy balance for routine deployment.

Transfer learning is not required in our present setting because the proposed network is lightweight and converges quickly.  Therefore, we train each rock model from scratch, which simplifies the pipeline and avoids dependence on pretrained weights from other rocks.
{Nevertheless, transfer learning remains a valid option} if only a small additional dataset is available for a new rock, since initializing the network weights from a previously trained rock model may reduce the amount of new training data required.


\section{Conclusions\label{sec:conclusion}}


We presented a fast surrogate workflow that combines a lightweight 3D CNN with hierarchical homogenization (HHM) to predict effective elastic properties of large digital rocks from segmented micro-CT data.
{In the proposed pipeline, each digital rock (here, a $900\times900\times900$ voxel volume) is partitioned into $75\times75\times75$ subcubes. Subcube-level elasticity is learned with one shared CNN backbone under three alternative targets: the full stiffness tensor (Model~1), direct $(K,G)$ (Model~2), or HS-normalized factors (Model~3). The resulting predictions are then embedded into HHM. Large-volume validation uses $300\times300\times300$ and $600\times600\times600$ crops extracted from within each parent rock, with CNN--HHM predictions compared against DNS on those same cropped regions.}
We evaluated this framework across five rock types (B1, B2, CG, FB1, and FB2) and assessed both subcube-level fidelity (parity/MSE and training cost) and HHM-level agreement with DNS on the cropped regions.

Across subcube-level and HHM-level evaluations, all three models achieved strong agreement with DNS references on B1, while substantially reducing the cost of the first HHM stage.
{The same trends hold for the other four rock types. At the $300\times300\times300$ scale, Models~2 and~3 consistently achieve higher $R^2$ and lower RMSE than Benchmark~1 (Tables~}\ref{tab:r2_300}{--}\ref{tab:rmse_300}{). In accuracy, Model~2 ranks first in most cases, with Model~3 close behind. Among our three models, Model~3 trains the fastest (Figure~}\ref{fig:eval_B1_cnn_training}{(c)).}
{For deployment, we recommend Model~3 for the best speed--accuracy balance and physics-consistent HS normalization. Model~2 is preferred when direct modulus outputs are desired. Model~1 remains the appropriate choice when there is reason to believe that the rock is elastically anisotropic at the subcube level and larger length scales.}
This recommendation is supported by our finding that, at least for B1, the isotropic-subcube assumption introduces little additional error.

\section*{Conflict of Interest}
The authors declare no competing interests.

\section*{Acknowledgments}

We acknowledge Shell for financial support and for providing the digital rock images.
%
This work uses computational resources provided by the Stanford Research Computing Center.
The authors would also like to thank Math2Market for providing the GeoDict software at a discount and for technical support.

\section*{Data availability}\label{sec:data_availability}

Related codes and data will be released via \url{https://gitlab.com/micronano_public/Digital_Rocks/CNN-HHM}

\appendix

\setcounter{figure}{0}
\setcounter{table}{0}
\setcounter{equation}{0}
\counterwithin{figure}{section}
\counterwithin{table}{section}
\counterwithin{equation}{section}
\renewcommand{\thefigure}{\thesection\arabic{figure}}
\renewcommand{\thetable}{\thesection\arabic{table}}
\renewcommand{\theequation}{\thesection\arabic{equation}}

\section{Stiffness matrix calculation using isotropic assumption}

When the CNN is trained to predict $(K,G)$ (or HS-normalized factors that are converted to $(K,G)$), we reconstruct an 
\emph{isotropic} stiffness tensor for use inside HHM.
For completeness, we summarize the mapping here.



The isotropic elastic stiffness tensor in ($6\times6$) Voigt form is
\[
\bm{\mathcal{C}} =
\begin{pmatrix}
\mathcal{C}_{11} & \mathcal{C}_{12} & \mathcal{C}_{12} & 0      & 0      & 0      \\[6pt]
\mathcal{C}_{12} & \mathcal{C}_{11} & \mathcal{C}_{12} & 0      & 0      & 0      \\[6pt]
\mathcal{C}_{12} & \mathcal{C}_{12} & \mathcal{C}_{11} & 0      & 0      & 0      \\[6pt]
0      & 0      & 0      & \mathcal{C}_{44} & 0      & 0      \\[6pt]
0      & 0      & 0      & 0      & \mathcal{C}_{44} & 0      \\[6pt]
0      & 0      & 0      & 0      & 0      & \mathcal{C}_{44} 
\end{pmatrix}.
\]
where
%
\[
\mathcal{C}_{11} = \lambda + 2\,G,\quad
\mathcal{C}_{12} = \lambda,\quad
\mathcal{C}_{44} = G.
\]
and
\(\lambda = K - \tfrac{2\,G}{3}\)
is the Lam\'e constant.
Therefore,
\begin{equation}
    \bm{\mathcal{C}} =
\begin{pmatrix}
K + \dfrac{4\,G}{3}   & K - \dfrac{2\,G}{3} & K - \dfrac{2\,G}{3} & 0 & 0 & 0 \\[8pt]
K - \dfrac{2\,G}{3}   & K + \dfrac{4\,G}{3} & K - \dfrac{2\,G}{3} & 0 & 0 & 0 \\[8pt]
K - \dfrac{2\,G}{3}   & K - \dfrac{2\,G}{3} & K + \dfrac{4\,G}{3} & 0 & 0 & 0 \\[8pt]
0                     & 0                   & 0                   & G & 0 & 0 \\[6pt]
0                     & 0                   & 0                   & 0 & G & 0 \\[6pt]
0                     & 0                   & 0                   & 0 & 0 & G
\end{pmatrix}.\label{eq:stiffness_isotropic}
\end{equation}


\section{Details on training CNN}

\subsection{CNN3D architecture details}

\begin{table}[htbp]
\centering
\caption{Architecture of the CNN3D surrogate model for elastic moduli prediction.}
\label{tab:cnn3d_arch}
\begin{tabular}{lcccc}
\toprule
\multicolumn{5}{c}{\textbf{The convolutional part}}\\
\midrule
Layer & Input size & Kernel size & \# Filters & Output size \\
\midrule
Conv3D (conv1)   & $1\times100\times100\times100$ & 3 & 8   & $8\times100\times100\times100$ \\
MaxPool3D (pool1)& $8\times100\times100\times100$ & 2 & --  & $8\times50\times50\times50$    \\
Conv3D (conv2)   & $8\times50\times50\times50$     & 3 & 32  & $32\times50\times50\times50$   \\
MaxPool3D (pool2)& $32\times50\times50\times50$   & 2 & --  & $32\times25\times25\times25$  \\
Conv3D (conv3)   & $32\times25\times25\times25$   & 3 & 128 & $128\times25\times25\times25$ \\
MaxPool3D (pool3)& $128\times25\times25\times25$ & 2 & --  & $128\times12\times12\times12$ \\
\bottomrule
\end{tabular}

\vspace{3pt}

\begin{tabular}{lcc}
\toprule
\multicolumn{3}{c}{\textbf{The fully connected part}}\\
\midrule
Layer             & Input size         & Output size    \\
\midrule
FC1 + ReLU         & $1\times221184$    & $1\times256$   \\
FC2 (stiffness)    & $1\times256$       & $1\times36$    \\
FC3 (moduli/HS)    & $1\times256$       & $1\times2$     \\
\bottomrule
\end{tabular}
\end{table}

%
%
Table~\ref{tab:cnn3d_arch} reports the model architecture implemented in this paper, including the shared convolutional feature extractor and the fully connected output heads.
The three Conv3D$+$MaxPool blocks reduce the $75\times75\times75$ 
voxel input to a compact feature representation, which is then flattened and passed through a 256-neuron fully connected layer.
The final layer is selected according to the training target: the stiffness-tensor model generates a 36-component output, whereas the direct-moduli and HS-normalized models generate a two-component output for either $(K,G)$ or $(f_K,f_G)$.

\subsection{Training on B1 rocks}

\begin{figure}[htbp]
    \centering
    \includegraphics[width=.65\linewidth]{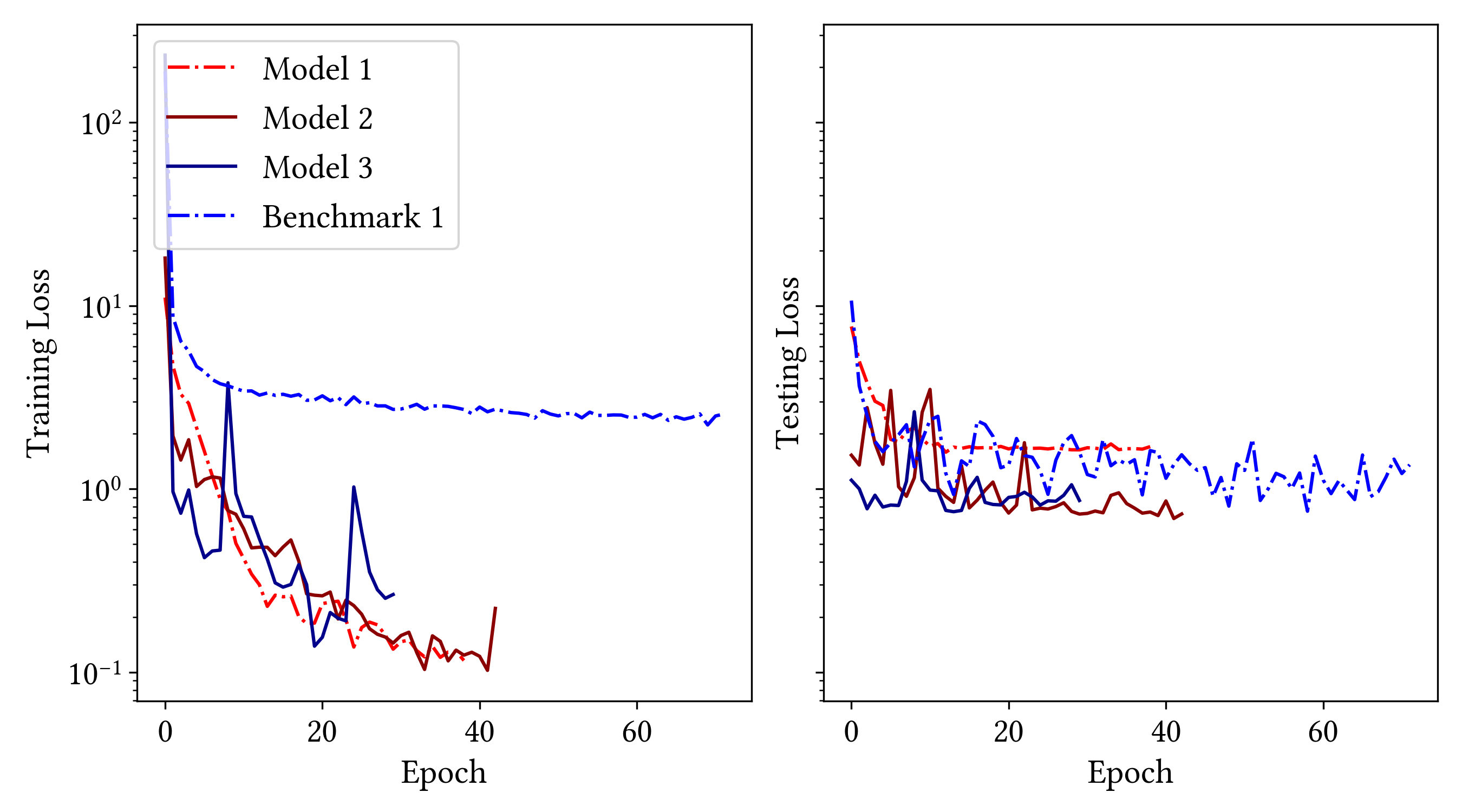}\\
    {\hspace{25pt}\centering(a)\hspace{155pt}(b)}
    \caption{Loss evolution during training on B1 subcubes.
    Solid and dashed curves show training and validation loss, respectively, for different CNN training targets.
    The plots show relative convergence behavior and the point at which early stopping terminates training.
    }
    \label{fig:loss_B1_learn}
\end{figure}

Figure~\ref{fig:loss_B1_learn} reports representative training and validation loss histories for four targets: HS-normalized factors ({Model 3}), HS-normalized factors ({Benchmark 1 from} Ref.~\cite{ahmad2023homogenizing}), elastic moduli $(K,G)$ ({Model 2}), and the full stiffness tensor ({Model 1}).
These curves illustrate convergence speed and the effect of early stopping across target choices.

\subsection{Prediction on additional rocks (\emph{ad hoc} training)}

\begin{figure}[htbp]
    \centering
    \includegraphics[width=\linewidth]{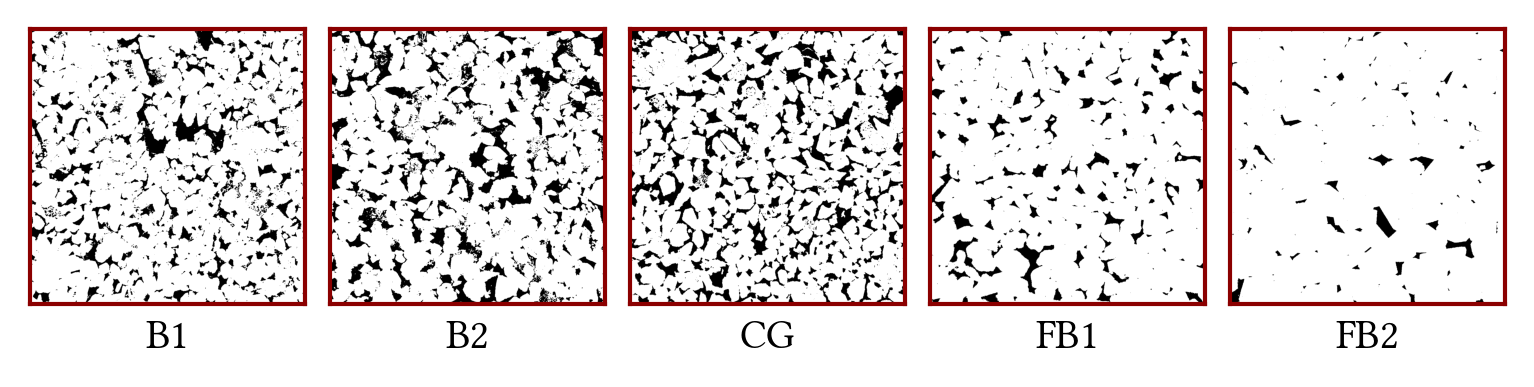}
    \caption{Representative 2D slices from the four additional rock types (B2, CG, FB1, FB2). The distinct pore morphologies and heterogeneity motivate evaluating model performance across rocks.}
    \label{fig:all_rock_slices}
\end{figure}

To put the cross-rock evaluation in context, Figure~\ref{fig:all_rock_slices} shows representative slices from each additional rock type (B1, B2, CG, FB1, FB2).
The visible differences in pore morphology and phase connectivity indicate that these rocks can exhibit distinct elastic responses.
We train CNNs \emph{ad hoc} on each rock type (rather than using transfer learning) and then apply the resulting CNN surrogates within HHM.
Tables~\ref{tab:r2_300} to \ref{tab:rmse_600}
below show that CNN+HHM yields effective moduli in good agreement with DNS on these additional rocks.

\begin{table*}[htbp]
\centering

\caption{$R^2$ values for the prediction of effective bulk ($K$) and shear ($G$) moduli using different modeling approaches on datasets with size-$300\times300\times300$ samples. Our Models 1, 2, 3 are defined in Section~\ref{sec:methods}.
Benchmark 1 is defined in Section~\ref{sec:results_discussion}.
Higher $R^2$ indicates better agreement with DNS results.}\label{tab:r2_300}
\begin{tabular}{l|cccc|cccc}
\hline
& \multicolumn{4}{c}{$K$} & \multicolumn{4}{c}{$G$} \\
\cline{2-5} \cline{6-9}
 
& Model 1 & Model 2 & Model 3 & Benchmark 1~\cite{ahmad2023homogenizing}
& Model 1 & Model 2 & Model 3 & Benchmark 1~\cite{ahmad2023homogenizing} \\
\hline

B1 & 0.966 & 0.988 & 0.993 & 0.957
   & 0.985 & 0.998 & 0.992 & 0.955 \\

B2 & 0.977 & 0.997 & 0.993 & 0.922
   & 0.982 & 0.998 & 0.995 & 0.912 \\

CG & 0.934 & 0.996 & 0.948 & 0.753
   & 0.942 & 0.997 & 0.967 & 0.761 \\

FB1 & 0.966 & 0.995 & 0.987 & 0.696
    & 0.972 & 0.996 & 0.991 & 0.691 \\

FB2 & 0.998 & 0.980 & 0.963 & 0.892
    & 0.997 & 0.990 & 0.970 & 0.878 \\

\hline
\end{tabular}
\end{table*}

\begin{table*}[htbp]
\centering

\caption{$R^2$ values for the prediction of effective bulk ($K$) and shear ($G$) moduli on datasets with size-$600\times600\times600$ samples. The same modeling approaches as in Table~1 are compared. The degradation of performance for some models and datasets highlights the sensitivity to increased variability and data complexity.}\label{tab:r2_600}
\begin{tabular}{l|cccc|cccc}
\hline
& \multicolumn{4}{c}{$K$} & \multicolumn{4}{c}{$G$} \\
\cline{2-5} \cline{6-9}
 
& Model 1 & Model 2 & Model 3 & Benchmark 1~\cite{ahmad2023homogenizing} 
& Model 1 & Model 2 & Model 3 & Benchmark 1~\cite{ahmad2023homogenizing} \\
\hline

B1 & 0.847 & 0.946 & 0.962 & 0.773
   & 0.901 & 0.995 & 0.955 & 0.720 \\

B2 & 0.837 & 0.997 & 0.971 & 0.481
   & 0.865 & 0.997 & 0.982 & 0.343 \\

CG & 0.673 & 0.991 & 0.750 & -0.306
   & 0.676 & 0.990 & 0.816 & -0.527 \\

FB1 & 0.720 & 0.971 & 0.919 & -1.608
    & 0.744 & 0.969 & 0.947 & -1.924 \\

FB2 & 0.990 & 0.752 & 0.543 & -0.204
    & 0.964 & 0.876 & 0.643 & -0.419 \\

\hline
\end{tabular}
\end{table*}

\begin{table*}[htbp]
\centering

\caption{Root mean square error (RMSE, in GPa) for the prediction of effective bulk ($K$) and shear ($G$) moduli using datasets with size-$300\times300\times300$ samples. Lower RMSE indicates higher predictive accuracy. Model definitions follow Section~\ref{sec:methods}.}\label{tab:rmse_300}
\begin{tabular}{l|cccc|cccc}
\hline
& \multicolumn{4}{c}{$K$} & \multicolumn{4}{c}{$G$} \\
\cline{2-5} \cline{6-9}
 
& Model 1 & Model 2 & Model 3 & Benchmark 1~\cite{ahmad2023homogenizing}
& Model 1 & Model 2 & Model 3 & Benchmark 1~\cite{ahmad2023homogenizing} \\
\hline

B1 & 0.32 & 0.19 & 0.15 & 0.36
   & 0.29 & 0.10 & 0.21 & 0.50 \\

B2 & 0.21 & 0.07 & 0.11 & 0.38
   & 0.24 & 0.08 & 0.13 & 0.52 \\

CG & 0.25 & 0.06 & 0.22 & 0.48
   & 0.30 & 0.07 & 0.23 & 0.61 \\

FB1 & 0.17 & 0.07 & 0.11 & 0.51
    & 0.22 & 0.09 & 0.12 & 0.74 \\

FB2 & 0.03 & 0.11 & 0.15 & 0.25
    & 0.07 & 0.12 & 0.20 & 0.40 \\

\hline
\end{tabular}
\end{table*}

\begin{table*}[htbp]
\centering

\caption{Root mean square error (RMSE, in GPa) for datasets with size-$600\times600\times600$ samples. The results demonstrate the robustness of the homogenized moduli and HS-based approaches compared to the reference model, particularly under increased dataset complexity.}\label{tab:rmse_600}
\begin{tabular}{l|cccc|cccc}
\hline
& \multicolumn{4}{c}{$K$} & \multicolumn{4}{c}{$G$} \\
\cline{2-5} \cline{6-9}
 
& Model 1 & Model 2 & Model 3 & Benchmark 1~\cite{ahmad2023homogenizing}
& Model 1 & Model 2 & Model 3 & Benchmark 1~\cite{ahmad2023homogenizing} \\
\hline

B1 & 0.26 & 0.16 & 0.13 & 0.32
   & 0.28 & 0.07 & 0.19 & 0.47 \\

B2 & 0.20 & 0.03 & 0.08 & 0.35
   & 0.24 & 0.04 & 0.09 & 0.52 \\

CG & 0.25 & 0.04 & 0.21 & 0.49
   & 0.29 & 0.05 & 0.22 & 0.63 \\

FB1 & 0.17 & 0.05 & 0.09 & 0.51
    & 0.22 & 0.08 & 0.10 & 0.75 \\

FB2 & 0.02 & 0.11 & 0.15 & 0.24
    & 0.06 & 0.12 & 0.20 & 0.39 \\

\hline
\end{tabular}
\end{table*}

The results in Tables~\ref{tab:r2_300}--\ref{tab:rmse_600} provide a comprehensive comparison of the predictive performance of different modeling approaches for estimating the effective bulk ($K$) and shear ($G$) moduli. Across all datasets, Model~2 (homogenized moduli) consistently achieves the highest $R^2$ values (Tables~\ref{tab:r2_300} and \ref{tab:r2_600}) and the lowest RMSE (Tables~\ref{tab:rmse_300} and \ref{tab:rmse_600}), indicating superior agreement with DNS results. 
Model~3 (HS bound factor) also performs robustly, particularly for the bulk modulus, where it often matches or closely follows Model~2.
The only exception occurs at the FB2 rock of size $600\times600\times600$, where the low $R^2$ value may be attributed to an insufficient number of testing data.
%
%
On the other hand, Model~1 (stiffness matrix) performs worse for CG and FB1 rocks, especially for the size-$600\times600\times600$ samples (Table~\ref{tab:r2_600}). 
%
Across the board, Models 1, 2, and 3 all exhibit higher accuracy than Benchmark 1~\cite{ahmad2023homogenizing} (greater $R^2$ values and lower RMSE). 
%

\section{Porosity effects on error}

\begin{figure}[h]
    \centering
    \includegraphics[width=0.55\textwidth]{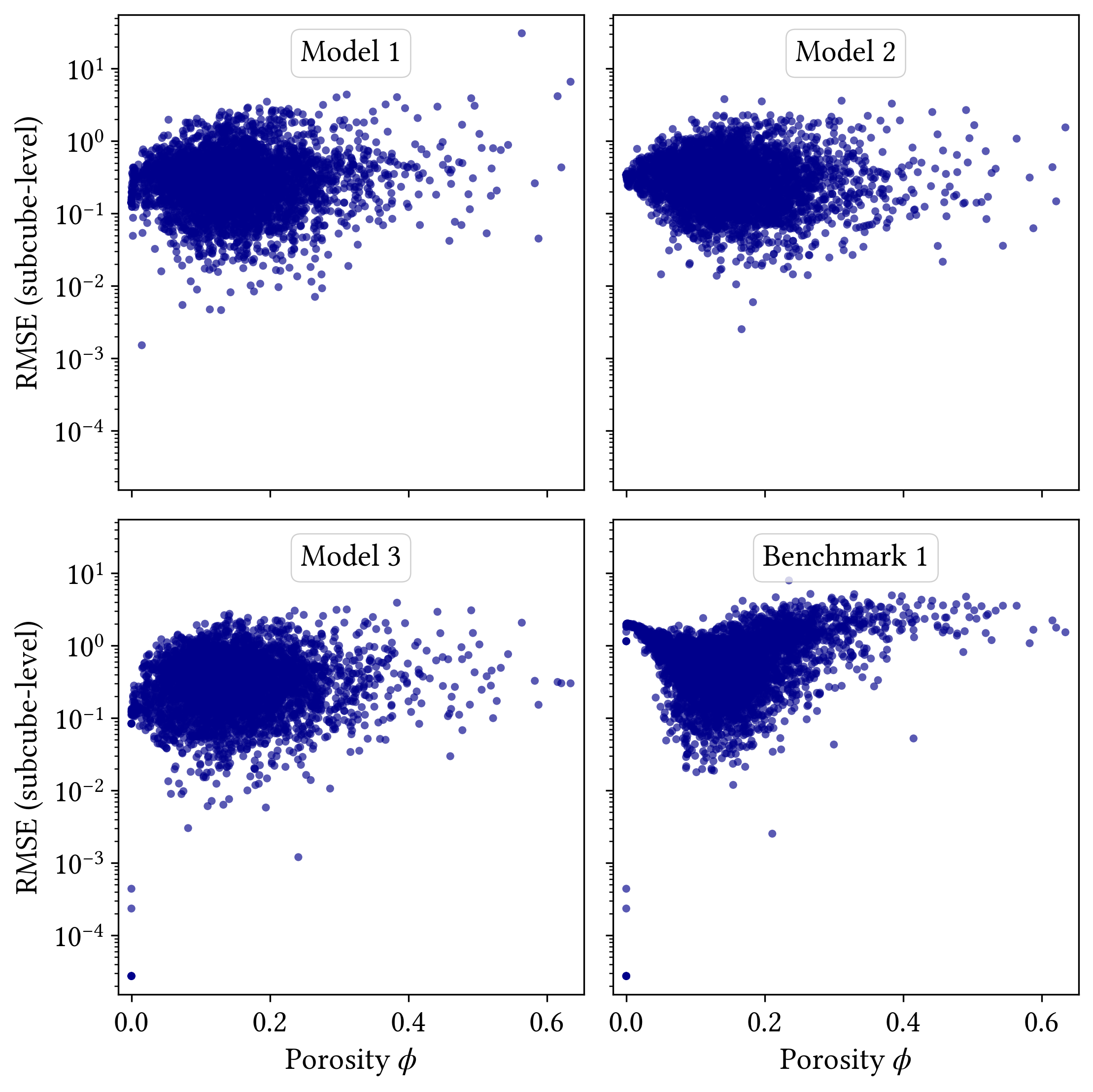}
    \caption{RMSE versus porosity for the B1 test set, comparing the three proposed CNN models and the reference model from Benchmark 1~\cite{ahmad2023homogenizing}.
    }
    \label{fig:phi_rmse_b1}
    
\end{figure}

It may be of interest to evaluate whether the error of the trained neural network models depend on the porosity of the subcubes.
For example, in the limit of zero porosity, the moduli of the subcube coincide with those of the mineral phase.
In principle, this should be an easy target for the neural network to learn.
Correspondingly, the upper and lower HS bounds coincide in the limit of zero porosity; hence our Model 3 is expected to recover the exact result by construction in this limit.

Figure~\ref{fig:phi_rmse_b1} {shows RMSE as a function of porosity for all three models and Benchmark 1.
In most cases, no clear trend is observed between error and porosity, suggesting that prediction accuracy is not strongly controlled by this microstructural descriptor.
{The only exceptions are in the three data points for porosity} $\phi < 1\times10^{-5}$, where both our Model 3 and Benchmark 1 show exceptionally low errors.
These results suggest that the porosity does not have a strong effect on the accuracy of the trained neural networks except when the porosity is extremely small.

\bibliographystyle{unsrt}
\bibliography{ref_cnn_paper}

@article{ahmad2023homogenizing,
  title={Homogenizing elastic properties of large digital rock images by combining CNN with hierarchical homogenization method},
  author={Ahmad, Rasool and Liu, Mingliang and Ortiz, Michael and Mukerji, Tapan and Cai, Wei},
  journal={arXiv preprint arXiv:2305.06519},
  year={2023},
  url={https://arxiv.org/abs/2305.06519}
}

@article{Tang2022_cnn_permeability,
  title = {Predicting permeability from 3D rock images based on CNN with physical information},
  volume = {606},
  ISSN = {0022-1694},
  url = {http://dx.doi.org/10.1016/j.jhydrol.2022.127473},
  DOI = {10.1016/j.jhydrol.2022.127473},
  journal = {Journal of Hydrology},
  publisher = {Elsevier BV},
  author = {Tang,  Pengfei and Zhang,  Dongxiao and Li,  Heng},
  year = {2022},
  month = mar,
  pages = {127473}
}

@article{Khan2024_cnn_permeability,
  title = {Machine Learning Assisted Prediction of Porosity and Related Properties Using Digital Rock Images},
  volume = {9},
  ISSN = {2470-1343},
  url = {http://dx.doi.org/10.1021/acsomega.3c10131},
  DOI = {10.1021/acsomega.3c10131},
  number = {28},
  journal = {ACS Omega},
  publisher = {American Chemical Society (ACS)},
  author = {Khan,  Md Irfan and Khanal,  Aaditya},
  year = {2024},
  month = jun,
  pages = {30205–30223}
}

@article{Bizhani2022_cnn_image_reconstruction,
  title = {Reconstructing high fidelity digital rock images using deep convolutional neural networks},
  volume = {12},
  ISSN = {2045-2322},
  url = {http://dx.doi.org/10.1038/s41598-022-08170-8},
  DOI = {10.1038/s41598-022-08170-8},
  number = {1},
  journal = {Scientific Reports},
  publisher = {Springer Science and Business Media LLC},
  author = {Bizhani,  Majid and Ardakani,  Omid Haeri and Little,  Edward},
  year = {2022},
  month = mar 
}

@inproceedings{deFigueiredo2019_cnn_p_velo,
  title = {Deep 3D convolutional neural network applied to CT segmented image for rock properties prediction},
  url = {http://dx.doi.org/10.1190/segam2019-3216835.1},
  DOI = {10.1190/segam2019-3216835.1},
  booktitle = {SEG Technical Program Expanded Abstracts 2019},
  publisher = {Society of Exploration Geophysicists},
  author = {de Figueiredo,  Leandro Passos and Bordignon,  Fernando and Exterkoetter,  Rodrigo and Rodrigues,  Bruno Barbosa and Duarte,  Maury},
  year = {2019},
  month = aug,
  pages = {xxx–xxx}
}

@article{Zheng2022_cnn_reconstruction,
  title = {Digital Rock Reconstruction with User-Defined Properties Using Conditional Generative Adversarial Networks},
  volume = {144},
  ISSN = {1573-1634},
  url = {http://dx.doi.org/10.1007/s11242-021-01728-6},
  DOI = {10.1007/s11242-021-01728-6},
  number = {1},
  journal = {Transport in Porous Media},
  publisher = {Springer Science and Business Media LLC},
  author = {Zheng,  Qiang and Zhang,  Dongxiao},
  year = {2022},
  month = jan,
  pages = {255–281}
}

@article{Hashin1963,
  title = {A variational approach to the theory of the elastic behaviour of multiphase materials},
  volume = {11},
  ISSN = {0022-5096},
  url = {http://dx.doi.org/10.1016/0022-5096(63)90060-7},
  DOI = {10.1016/0022-5096(63)90060-7},
  number = {2},
  journal = {Journal of the Mechanics and Physics of Solids},
  publisher = {Elsevier BV},
  author = {Hashin,  Z. and Shtrikman,  S.},
  year = {1963},
  month = mar,
  pages = {127–140}
}

@article{Schepp2020Digital,
  author  = {Schepp, Laura L. and Ahrens, Benedikt and Balcewicz, Martin and Duda, Mandy and Nehler, Mathias and Osorno, Maria and Uribe, David and Steeb, Holger and Nigon, Benoit and Stöckhert, Ferdinand and Swanson, Donald A. and Siegert, Mirko and Gurris, Marcel G. and Saenger, Erik H.},
  title   = {Digital rock physics and laboratory considerations on a high-porosity volcanic rock},
  journal = {Scientific Reports},
  volume  = {10},
  pages   = {5840},
  year    = {2020},
  doi     = {10.1038/s41598-020-62741-1},
}

@article{Lin2019Validation,
  author  = {Lin, Rongrong and Thomsen, Leon},
  title   = {Validation of Digital Rock Physics Algorithms},
  journal = {Minerals},
  volume  = {9},
  number  = {11},
  pages   = {669},
  year    = {2019},
  doi     = {10.3390/min9110669},
}

@article{Goldfarb2022Predictive,
  author  = {Goldfarb, Eric J. and Ikeda, Ken and Ketcham, Richard A. and Spikes, Kyle T. and Prodanovi{\'c}, Ma{\v{s}}a and Tisato, Nicola},
  title   = {Predictive digital rock physics without segmentation},
  journal = {Computers \& Geosciences},
  volume  = {159},
  pages   = {105008},
  year    = {2022},
  doi     = {10.1016/j.cageo.2021.105008},
}

@article{Ahmad2022,
  author    = {Ahmad, Rasool and Liu, Mingliang and Ortiz, Michael and Mukerji, Tapan and Cai, Wei},
  title     = {Computation of effective elastic moduli of rocks using hierarchical homogenization},
  journal   = {Journal of the Mechanics and Physics of Solids},
  doi       = {10.1016/j.jmps.2023.105268},
  year      = {2023},
}

@article{Andr2013_drp_1,
  title = {Digital rock physics benchmarks—Part I: Imaging and segmentation},
  volume = {50},
  ISSN = {0098-3004},
  url = {http://dx.doi.org/10.1016/j.cageo.2012.09.005},
  DOI = {10.1016/j.cageo.2012.09.005},
  journal = {Computers \& Geosciences},
  publisher = {Elsevier BV},
  author = {Andr\"{a},  Heiko and Combaret,  Nicolas and Dvorkin,  Jack and Glatt,  Erik and Han,  Junehee and Kabel,  Matthias and Keehm,  Youngseuk and Krzikalla,  Fabian and Lee,  Minhui and Madonna,  Claudio and Marsh,  Mike and Mukerji,  Tapan and Saenger,  Erik H. and Sain,  Ratnanabha and Saxena,  Nishank and Ricker,  Sarah and Wiegmann,  Andreas and Zhan,  Xin},
  year = {2013},
  month = jan,
  pages = {25–32}
}

@article{Andr2013_drp_2,
  title = {Digital rock physics benchmarks—part II: Computing effective properties},
  volume = {50},
  ISSN = {0098-3004},
  url = {http://dx.doi.org/10.1016/j.cageo.2012.09.008},
  DOI = {10.1016/j.cageo.2012.09.008},
  journal = {Computers \& Geosciences},
  publisher = {Elsevier BV},
  author = {Andr\"{a},  Heiko and Combaret,  Nicolas and Dvorkin,  Jack and Glatt,  Erik and Han,  Junehee and Kabel,  Matthias and Keehm,  Youngseuk and Krzikalla,  Fabian and Lee,  Minhui and Madonna,  Claudio and Marsh,  Mike and Mukerji,  Tapan and Saenger,  Erik H. and Sain,  Ratnanabha and Saxena,  Nishank and Ricker,  Sarah and Wiegmann,  Andreas and Zhan,  Xin},
  year = {2013},
  month = jan,
  pages = {33–43}
}

@book{voigt1928kristallphysik,
  author    = {Voigt, Woldemar},
  title     = {Lehrbuch der Kristallphysik},
  year      = {1928},
  publisher = {Teubner Verlag},
  address   = {Leipzig}
}

@article{reuss1929,
  author  = {Reuss, A.},
  title   = {Berechnung der Fließgrenze von Mischkristallen auf Grund der Plastizitätsbedingung für Einkristalle},
  journal = {Zeitschrift für Angewandte Mathematik und Mechanik},
  volume  = {9},
  number  = {1},
  pages   = {49--58},
  year    = {1929},
  doi     = {10.1002/zamm.19290090104}
}

@article{hill1952,
  author  = {Hill, R.},
  title   = {The Elastic Behaviour of a Crystalline Aggregate},
  journal = {Proceedings of the Physical Society. Section A},
  volume  = {65},
  number  = {5},
  pages   = {349--354},
  year    = {1952},
  doi     = {10.1088/0370-1298/65/5/307}
}

@article{Elmorsy2022_CNN_permeability,
  title = {Generalizable Permeability Prediction of Digital Porous Media via a Novel Multi‐Scale 3D Convolutional Neural Network},
  volume = {58},
  ISSN = {1944-7973},
  url = {http://dx.doi.org/10.1029/2021WR031454},
  DOI = {10.1029/2021wr031454},
  number = {3},
  journal = {Water Resources Research},
  publisher = {American Geophysical Union (AGU)},
  author = {Elmorsy,  Mohamed and El‐Dakhakhni,  Wael and Zhao,  Benzhong},
  year = {2022},
  month = mar 
}

@article{Karimpouli2019_cnn_s_velo,
  title = {Image-based velocity estimation of rock using Convolutional Neural Networks},
  volume = {111},
  ISSN = {0893-6080},
  url = {http://dx.doi.org/10.1016/j.neunet.2018.12.006},
  DOI = {10.1016/j.neunet.2018.12.006},
  journal = {Neural Networks},
  publisher = {Elsevier BV},
  author = {Karimpouli,  Sadegh and Tahmasebi,  Pejman},
  year = {2019},
  month = mar,
  pages = {89–97}
}

@article{Moulinec1998_fft_solver,
  title = {A numerical method for computing the overall response of nonlinear composites with complex microstructure},
  volume = {157},
  ISSN = {0045-7825},
  url = {http://dx.doi.org/10.1016/S0045-7825(97)00218-1},
  DOI = {10.1016/s0045-7825(97)00218-1},
  number = {1–2},
  journal = {Computer Methods in Applied Mechanics and Engineering},
  publisher = {Elsevier BV},
  author = {Moulinec,  H. and Suquet,  P.},
  year = {1998},
  month = apr,
  pages = {69–94}
}

@article{Hou2022_moduli_cnn,
  title = {Estimating elastic parameters from digital rock images based on multi-task learning with multi-gate mixture-of-experts},
  volume = {213},
  ISSN = {0920-4105},
  url = {http://dx.doi.org/10.1016/j.petrol.2022.110310},
  DOI = {10.1016/j.petrol.2022.110310},
  journal = {Journal of Petroleum Science and Engineering},
  publisher = {Elsevier BV},
  author = {Hou,  Zhiyu and Cao,  Danping},
  year = {2022},
  month = jun,
  pages = {110310}
}

@article{Zorkaltsev2025_moduli_cnn,
  title = {Transferable 3D convolutional neural networks for elastic constants prediction in nanoporous metals},
  volume = {260},
  ISSN = {0264-1275},
  url = {http://dx.doi.org/10.1016/j.matdes.2025.114896},
  DOI = {10.1016/j.matdes.2025.114896},
  journal = {Materials \& Design},
  publisher = {Elsevier BV},
  author = {Zorkaltsev,  Sergei and Topolnicki,  Rafał and Carmon,  Tal-El and Mathesan,  Santhosh and Dłotko,  Paweł and Mordehai,  Dan and Haranczyk,  Maciej},
  year = {2025},
  month = dec,
  pages = {114896}
}

@misc{GeoDict_software,
  title        = {GeoDict: Digital Material Research Software},
  author       = {{Math2Market GmbH}},
  year         = {2023},
  howpublished = {Software},
  url          = {https://www.math2market.com/geodict-software}
}

@article{nishank_mech_estimation,
  title = {Effect of image segmentation \& voxel size on micro-CT computed effective transport \& elastic properties},
  volume = {86},
  ISSN = {0264-8172},
  url = {http://dx.doi.org/10.1016/j.marpetgeo.2017.07.004},
  DOI = {10.1016/j.marpetgeo.2017.07.004},
  journal = {Marine and Petroleum Geology},
  publisher = {Elsevier BV},
  author = {Saxena,  Nishank and Hofmann,  Ronny and Alpak,  Faruk O. and Dietderich,  Jesse and Hunter,  Sander and Day-Stirrat,  Ruarri J.},
  year = {2017},
  month = sep,
  pages = {972–990}
}

@article{hashin1962some,
  title={On some variational principles in anisotropic and nonhomogeneous elasticity},
  author={Hashin, Z am and Shtrikman, SJJotM},
  journal={Journal of the Mechanics and Physics of Solids},
  volume={10},
  number={4},
  pages={335--342},
  year={1962},
  publisher={Elsevier}
}
\end{document}